\documentclass[paper,notoc,nohyper]{JHEP}
\usepackage{epsfig}
\def\S{S_{\epsilon}}

\def\F{{\cal F}}

\def\Bfin{{\rm Box^6} (u, t) }
\def\Bubl{{\rm Bub}(s)}

\def\lib{{\rm Li}_2}
\def\lic{{\rm Li}_3}
\def\lid{{\rm Li}_4}
\def\Ls{L_s}
\def\lnx{L_x}
\def\lny{L_y}
\def\Poles{{\cal P}oles}
\def\Finite{{\cal F}inite}
\def\Libx{{\rm Li}_2(x)}
\def\Liby{{\rm Li}_2(y)}
\def\Licx{{\rm Li}_3(x)}
\def\Licy{{\rm Li}_3(y)}
\def\Lidx{{\rm Li}_4(x)}
\def\Lidy{{\rm Li}_4(y)}
\def\Lidz{{\rm Li}_4\Biggl(\frac{x-1}{x}\Biggr)}

\def\tsmus{\left[\frac{t^2-u^2}{s^2}\right]}
\def\ssmus{\left[\frac{s^2-u^2}{t^2}\right]}
\def\tsmss{\left[\frac{t^2-s^2}{u^2}\right]}
\def\tspus{\left[\frac{t^2+u^2}{s^2}\right]}
\def\sspus{\left[\frac{s^2+u^2}{t^2}\right]}
\def\tspss{\left[\frac{t^2+s^2}{u^2}\right]}

\def\tou{\frac{t^3}{s^2u}}
\def\uot{\frac{u^3}{s^2t}}
\def\sou{\frac{s^3}{t^2u}}
\def\uos{\frac{u^3}{t^2s}}
\def\tos{\frac{t^3}{u^2s}}
\def\sot{\frac{s^3}{u^2t}}

\def\CA{C_A}
\def\CF{C_F}

\def\A{{\cal A}}
\renewcommand\O[1]{{\cal O}\left(#1\right)}
\def\as{\ensuremath{\alpha_{s}}}

\def\Re{\mathop{\rm Re}}

\def\t3ou{\,\frac{t^3}{us^2}}
\def\u3ot{\,\frac{u^3}{ts^2}}

\def\NF{\,N_F}

\def\beq{\begin{equation}}
\def\eeq{\end{equation}}

\def\beqn{\begin{eqnarray}}
\def\eeqn{\end{eqnarray}}

\def\lq{\left[}
\def\rq{\right]}

\def\({\left(}
\def\){\right)}

\def\ket#1{|{#1}\rangle}
\def\bra#1{\langle{#1}|}
\def\braket#1#2{\langle #1 |#2 \rangle}
\def\cm{{\cal M}}
\def\t#1#2#3{t^#1_{#2\,#3}}

\def\MSbar{$\overline{{\rm MS}}$}

\def\a0{\alpha_0}

\def\ab0{\bar{\alpha}_0}

\def\bom#1{{\mbox{\boldmath $#1$}}}

\def\TTT#1#2{ {\bom T}_#1 \cdot {\bom T}_#2 }

\def\fs{\(-\frac{\mu^2}{s}\)^\ep }
\def\ft{\(-\frac{\mu^2}{t}\)^\ep }
\def\fu{\(-\frac{\mu^2}{u}\)^\ep }
\def\fsd{\(-\frac{\mu^2}{s}\)^{2\ep} }
\def\ftd{\(-\frac{\mu^2}{t}\)^{2\ep} }
\def\fud{\(-\frac{\mu^2}{u}\)^{2\ep} }
\def \ep{\epsilon}
\def\ord#1{{\cal O}\(#1\)}

\newcommand{\SUNC}[1]{
\mbox{\parbox{3cm}{
\begin{picture}(3.5,1.4)
\thicklines
\put(0.5,0.7){\line(1,0){2}}
\put(1.5,0.7){\circle{1}}
\put(2.65,0.7){\makebox(0,0)[l]{$(#1)$}}
\end{picture}
}}
\hfill}
\newcommand{\BUB}[1]{
\mbox{\parbox{3cm}{
\begin{picture}(3.5,1.4)
\thicklines
\put(0.5,0.7){\line(1,0){0.5}}
\put(2.0,0.7){\line(1,0){0.5}}
\put(1.5,0.7){\circle{1}}
\put(2.65,0.7){\makebox(0,0)[l]{$(#1)$}}
\end{picture}
}}
\hfill}

\newcommand{\GLASS}[1]{
\mbox{\parbox{3cm}{
\begin{picture}(3.5,1.4)
\thicklines
\put(0.5,0.7){\line(1,0){0.5}}
\put(3,0.7){\line(1,0){0.5}}
\put(1.5,0.7){\circle{1}}
\put(2.5,0.7){\circle{1}}
\put(3.65,0.7){\makebox(0,0)[l]{$(#1)$}}
\end{picture}
}}
\hfill}

\newcommand{\TRI}[1]{
\mbox{\parbox{3cm}{
\begin{picture}(3.5,1.4)
\thicklines
\put(0.5,0.7){\line(1,0){0.5}}
\put(1.5,1.2){\line(0,-1){1}}
\put(1.5,1.2){\line(1,0){1}}
\put(1.5,0.2){\line(1,0){1}}
\put(1.5,0.7){\circle{1}}
\put(2.65,0.7){\makebox(0,0)[l]{$(#1)$}}
\end{picture}
}}
\hfill}

\newcommand{\ABOX}[2]{
\mbox{\parbox{3cm}{
\begin{picture}(3.5,1.4)
\thicklines
\put(0.5,0.2){\line(1,0){2.4}}
\put(0.5,1.2){\line(1,0){2.4}}
\put(1,0.2){\line(0,1){1}}
\put(2,0.7){\circle{1}}
\put(3,0.7){\makebox(0,0)[l]{$(#1,#2)$}}
\end{picture}
}}
\hfill}

\newcommand{\CBOX}[2]{
\mbox{\parbox{3cm}{
\begin{picture}(3.5,1.4)
\thicklines
\put(0.5,0.2){\line(1,0){2.4}}
\put(1.2,0.2){\line(1,1){1}}
\put(0.5,1.2){\line(1,0){2.4}}
\put(1.2,0.2){\line(0,1){1}}
\put(2.2,0.2){\line(0,1){1}}
\put(3.05,0.7){\makebox(0,0)[l]{$(#1,#2)$}}
\end{picture}
}} 
\hfill}
\newcommand{\BOX}[2]{
\mbox{\parbox{3cm}{
\begin{picture}(3.5,1.4)
\thicklines
\put(0.5,0.2){\line(1,0){2.4}}
\put(0.5,1.2){\line(1,0){2.4}}
\put(1.2,0.2){\line(0,1){1}}
\put(2.2,0.2){\line(0,1){1}}
\put(3.05,0.7){\makebox(0,0)[l]{$(#1,#2)$}}
\end{picture}
}} 
\hfill}

\newcommand{\Pboxa}[2]{
\mbox{\parbox{3cm}{
\begin{picture}(3.5,1.4)
\thicklines
\put(0.5,0.2){\line(1,0){2.4}}
\put(1.7,0.2){\line(0,1){1}}
\put(0.5,1.2){\line(1,0){2.4}}
\put(1,0.2){\line(0,1){1}}
\put(2.4,0.2){\line(0,1){1}}
\put(3.05,0.7){\makebox(0,0)[l]{$(#1,#2)$}}
\end{picture}
}} 
\hfill}

\newcommand{\Pboxb}[2]{
\mbox{\parbox{3cm}{
\begin{picture}(3.5,1.4)
\thicklines
\put(0.5,0.2){\line(1,0){2.4}}
\put(1.7,0.2){\line(0,1){1}}
\put(0.5,1.2){\line(1,0){2.4}}
\put(1,0.2){\line(0,1){1}}
\put(2.4,0.2){\line(0,1){1}}
\put(1.35,0.7){\circle{0.5}}
\put(1.35,0.7){\makebox(0,0)[c]{$1$}}
\put(3.05,0.7){\makebox(0,0)[l]{$(#1,#2)$}}
\end{picture}
}} 
\hfill}
\newcommand{\Pboxd}[2]{
\mbox{\parbox{3cm}{
\begin{picture}(3.5,1.4)
\thicklines
\put(0.5,0.2){\line(1,0){2.4}}
\put(1.7,0.2){\line(0,1){1}}
\put(0.5,1.2){\line(1,0){2.4}}
\put(1,0.2){\line(0,1){1}}
\put(2.4,0.2){\line(0,1){1}}
\put(2.05,0.7){\circle{0.5}}
\put(2.05,0.7){\makebox(0,0)[c]{$1$}}
\put(3.05,0.7){\makebox(0,0)[l]{$(#1,#2)$}}
\end{picture}
}} 
\hfill}

\newcommand{\Pboxc}[2]{
\mbox{\parbox{3cm}{
\begin{picture}(3.5,1.4)
\thicklines
\put(0.5,0.2){\line(1,0){2.4}}
\put(1.7,0.2){\line(0,1){1}}
\put(0.5,1.2){\line(1,0){2.4}}
\put(1,0.2){\line(0,1){1}}
\put(2.4,0.2){\line(0,1){1}}
\put(1.7,0.7){\circle*{0.2}}
\put(3.05,0.7){\makebox(0,0)[l]{$(#1,#2)$}}
\end{picture}
}} 
\hfill}

\newcommand{\Xboxa}[2]{
\mbox{\parbox{3cm}{
\begin{picture}(3.5,1.4)
\thicklines
\put(0.5,0.2){\line(1,0){2.4}}
\put(0.5,1.2){\line(1,0){2.4}}
\put(1,0.2){\line(0,1){1}}
\put(1.7,0.2){\line(1,1){1}}
\put(1.7,1.2){\line(1,-1){1}}
\put(3.05,0.7){\makebox(0,0)[l]{$(#1,#2)$}}
\end{picture}
}} 
\hfill}

\newcommand{\Xboxb}[2]{
\mbox{\parbox{3cm}{
\begin{picture}(3.5,1.4)
\thicklines
\put(0.5,0.2){\line(1,0){2.4}}
\put(1.7,0.2){\line(1,1){1}}
\put(0.5,1.2){\line(1,0){2.4}}
\put(1,0.2){\line(0,1){1}}
\put(1.7,1.2){\line(1,-1){1}}
\put(1,0.7){\circle*{0.2}}
\put(3.05,0.7){\makebox(0,0)[l]{$(#1,#2)$}}
\end{picture}
}} 
\hfill}

\newcommand{\Xtri}[1]{
\mbox{\parbox{3cm}{
\begin{picture}(3.5,1.4)
\thicklines
\put(0.5,0.7){\line(1,0){0.7}}
\put(1.2,0.7){\line(3,1){2.2}}
\put(1.2,0.7){\line(3,-1){2.2}}
\put(1.8,0.9){\line(3,-2){1.2}}
\put(1.8,0.5){\line(3,2){1.2}}
\put(3.05,0.7){\makebox(0,0)[l]{$(#1)$}}
\end{picture}
}} 
\hfill}

\title{\boldmath Two-loop QCD corrections to the scattering of 
massless distinct quarks\footnote{Work supported in part by the UK Particle Physics and
Astronomy Research Council and by the EU Fourth Framework Programme
`Training and Mobility of Researchers', Network `Quantum Chromodynamics
and the Deep Structure of Elementary Particles',
contract FMRX-CT98-0194 (DG 12 - MIHT).
C.A. acknowledges
the financial support of the Greek government and
M.E.T. acknowledges financial support
from CONACyT and the CVCP. We thank
the British Council and German Academic Exchange Service for support
under ARC project 1050.
}}
\author{
C.~Anastasiou$^a$,
E.~W.~N.~Glover$^a$,
C.~Oleari$^b$ and M.~E.~Tejeda-Yeomans$^a$\\
$^a$Department of Physics, 
University of Durham, 
Durham DH1 3LE, 
England\\[1mm]
$^b$Department of Physics, 
University of Wisconsin,
1150 University Avenue\\
Madison WI 53706, 
U.S.A.\\[1mm]
E-mail: \email{Ch.Anastasiou@durham.ac.uk}, \email{E.W.N.Glover@durham.ac.uk},
\email{Oleari@pheno.physics.wisc.edu}, 
\email{M.E.Tejeda-Yeomans@durham.ac.uk}}
\abstract{
We present the two-loop virtual QCD corrections to the scattering of distinct
massless quarks, $q \bar q \to 
q^\prime \bar{q}^\prime$, in conventional dimensional
regularisation.  The structure of the infrared divergences agrees with
that predicted by Catani while expressions for the finite
remainder are given for each of the $s$-, $t$- and $u$-channels in
terms of polylogarithms.   The results presented here form a vital
part of the next-to-next-to-leading order contribution to inclusive 
jet production in hadron colliders and will play a crucial
role in improving the theoretical prediction for jet cross 
sections in
hadron-hadron collisions. }
\preprint{DTP/00/66,{IPPP/00/05},
{MADPH-00-1197},{hep-ph/0010212}
}

\begin{document}

\section{Introduction}
\label{sec:intro}

In hadron-hadron collisions, the most basic hard process is parton-parton
scattering to form a large transverse momentum jet. The single jet inclusive
transverse energy distribution observed at the TEVATRON and CERN S$p\bar{p}$S
shows good agreement with theoretical next-to-leading order $\O{\as^3}$ perturbative
predictions over a wide range of jet transverse energies and tests the
point-like nature of the partons down to distance scales of $10^{-17}$~m.  
However,  data collected in Run I by the  CDF collaboration at the TEVATRON
indicated possible new physics at large transverse energy~\cite{CDF}.  Data
obtained by  the D0 collaboration~\cite{D0} was more consistent with
next-to-leading order expectations. However, because of both 
theoretical and experimental uncertainties no definite conclusion could be
drawn. The experimental situation may be clarified in the forthcoming 
Run II starting in 2001 where increased statistics and improved
detectors may lead to a reduction in both the statistical and systematic
errors.

The theoretical prediction may be improved by including the
next-to-next-to-leading order perturbative predictions.  This has the effect of
(a) reducing the renormalisation scale dependence and (b) improving the
matching of the parton level theoretical jet algorithm with the hadron level
experimental jet algorithm because the jet structure can be modeled by the
presence of a third parton. Varying the renormalisation scale up and down by a
factor of two about the jet transverse energy leads to a 20\% (10\%)
renormalisation scale uncertainty at leading order (next-to-leading order)
for jets with $E_T \sim 100$~GeV. The improvement in accuracy
expected at next-to-next-to-leading order can be estimated using the
renormalisation group equations together with the known  leading and
next-to-leading order coefficients and is at the 1-2\% level. 

The full next-to-next-to-leading order prediction  requires a
knowledge of the two-loop $2 \to 2$ matrix elements as well as the
contributions from the one-loop  $2 \to 3$ and tree-level $2 \to 4$
processes.   In the interesting
large-transverse-energy region, $E_T \gg m_{\rm quark}$, 
the quark masses may be
safely neglected and we therefore focus on the scattering of massless partons.
For processes involving up, down and strange quarks, which together with
processes involving gluons form the bulk of the cross section, 
this is certainly a reliable approximation.   
The contribution involving charm and bottom quarks 
is only a small part of the total since the parton densities for finding 
charm and bottom quarks inside the proton are relatively suppressed. 
We note that the existing next-to-leading order 
programs~\cite{EKS,jetrad} used to compare directly with the experimental 
jet data~\cite{CDF,D0} are based on
massless parton-parton scattering.
Helicity amplitudes for the one-loop $2 \to 3$ parton
sub-processes  $gg\to ggg$, $\bar{q}q\to ggg$,
$\bar{q}q\to\bar{q}^\prime q^\prime g$, and processes related to these by
crossing 
symmetry, have been computed in~\cite{5g,3g2q,1g4q} respectively. The
amplitudes for the six gluon $gg\to gggg$, four gluon-two quark 
$\bar{q}q\to gggg$, two gluon-four quark
$\bar{q}q\to\bar{q}^\prime q^\prime gg$ and six quark
$\bar{q}q\to\bar{q}^\prime q^\prime\bar{q}^{\prime\prime} q^{\prime\prime}$
$2\to 4$ processes and the associated crossed processes computed at
tree-level are also known and are available in
\cite{6g,4g2q,2g4q,6q}.  

The calculation of the two-loop amplitudes for the massless  
$2 \to 2$ scattering processes
\begin{eqnarray} 
\label{eq:qqQQ} 
q + \bar{q} & \to & q^\prime + \bar{q}^\prime \\ 
q + \bar{q} & \to & q + \bar{q},  \\ 
q + \bar{q} & \to & g + g, \\ 
g + g &\to & g + g, 
\end{eqnarray} 
has proved more intractable due mainly to the
difficulty of evaluating the planar and non-planar double box graphs. Recently
however, analytic expressions  for these basic scalar integrals for massless
particle scattering have been
provided by Smirnov~\cite{planarA} and by Tausk~\cite{nonplanarA}  as series 
in $\ep=(4-D)/2$.  
Associated tensor integrals have also been solved in~\cite{planarB}
and~\cite{nonplanarB} so that generic two-loop massless  $2 \to 2$ processes
can in principle be expressed in terms of a basis set of known two-loop
integrals. With the notable exception of the maximal helicity violating two
loop amplitude for $gg \to gg$ which has recently been calculated by Bern,
Dixon and Kosower~\cite{bdk}\footnote{This amplitude vanishes at tree level and
does therefore not contribute to $2 \to 2$ scattering at
next-to-next-to-leading order $\O{\as^4}$.}, the two-loop matrix elements for
the $2 \to 2$  QCD parton scattering processes are not known.  It is the purpose of
this paper to provide dimensionally regularised and renormalised
analytic expressions 
at the two-loop level for process
(\ref{eq:qqQQ}) together with the time-reversed and crossed processes
\begin{eqnarray*}
q + q^\prime &\to & q + q^\prime,\\
q + \bar{q}^\prime &\to & q + \bar{q}^\prime,\\
\bar{q} + \bar{q}^\prime &\to & \bar{q} + \bar{q}^\prime.
\end{eqnarray*}
As is common in QCD calculations, we use the \MSbar\  renormalisation
scheme and conventional dimensional regularisation  where all external
particles are
treated in $D$ dimensions. 
We note that Bern, Dixon and Ghinculov~\cite{BDG} have recently
completed the first full two-loop calculation of physical $2 \to 2$
scattering amplitudes, the QED  processes $e^+e^- \to \mu^+\mu^-$ and
$e^+e^- \to e^-e^+$.    There is an overlap between their QED
calculation and the QCD results presented here and we expect that 
the analytic expressions presented here  will therefore provide a
useful check of some of their results.

Our paper is structured as follows.  In Section~\ref{sec:notation} we
define our notation while a brief description of the methodology is
given in Section~\ref{sec:method}. The results are collected in
Section~\ref{sec:results} where we provide analytic expressions for
the interference of the two-loop and tree-level amplitudes as series
expansions in $\ep$.  Catani has described the pole structure of
generic renormalised two-loop amplitudes~\cite{catani} and we use his
techniques to isolate the poles in the \MSbar\  scheme.    We
find that the pole structure expected in the \MSbar\  scheme
on general grounds is indeed reproduced by direct evaluation of the
Feynman diagrams. Ultimately these poles must be canceled by
infrared singularities from tree level $2 \to 4$  and one-loop $2 \to
3$ processes.   The finite remainder of the two-loop graphs form the
main results of our paper and are given in Section~\ref{sec:results}.   
Our findings are summarized in Section~\ref{sec:conc}.

\section{Notation}
\label{sec:notation}

For calculational convenience, we treat all particles as incoming so that
\begin{equation}
\label{eq:proc}
q (p_1) + \bar q (p_2)  + q^\prime (p_3) + \bar{q}^\prime (p_4) \to 0
\end{equation}
where the light-like momentum assignments are in parentheses and satisfy
$$
p_1^\mu+p_2^\mu+p_3^\mu+p_4^\mu = 0.
$$

As stated above,
we work in conventional dimensional regularisation treating all external
states in $D$ dimensions.  We renormalise in the \MSbar\  scheme where
the bare coupling $\a0$ is related to the running coupling $\as \equiv \alpha_s(\mu^2)$  
at renormalisation scale $\mu$ via
\beq
\label{eq:alpha}
\a0 \, \S = \as \, \lq 1 - \frac{\beta_0}{\ep}  
\, \left(\frac{\as}{2\pi}\right) + \( \frac{\beta_0^2}{\ep^2} - \frac{\beta_1}{2\ep} \)  \, 
\left(\frac{\as}{2\pi}\right)^2
+\ord{\as^3} \rq.
\eeq
In this expression
\beq
\S = (4 \pi)^\ep e^{-\ep \gamma},  \quad\quad \gamma=0.5772\ldots=
{\rm Euler\ constant}
\eeq
is the typical phase-space volume factor in $D=4-2\ep$ dimensions, and
$\beta_0, \beta_1$ are the first two coefficients of the QCD beta function for $\NF$ 
(massless) quark flavours
\beq
\label{betas}
\beta_0 = \frac{11 \CA - 4 T_R \NF}{6} \;\;, \;\; \;\;\;\;
\beta_1 = \frac{17 \CA^2 - 10 \CA T_R \NF - 6 \CF T_R \NF}{6} \;\;.
\eeq
For an $SU(N)$ gauge theory, where $N$ is the number of colours
\beq
\CF = \left(\frac{N^2-1}{2N}\right), \qquad \CA = N, \qquad T_R = \frac{1}{2}.
\eeq
The renormalised four point amplitude in the \MSbar\  scheme is thus
\beq
\ket{\cm}= 4\pi \as \lq \ket{\cm^{(0)}} + \left(\frac{\as}{2\pi}\right) \,\ket{\cm^{(1)}}
+ \left(\frac{\as}{2\pi}\right)^2\,
\ket{ \cm^{(2)}}+ \ord{\as^3} \rq,
\eeq
where the $\ket{\cm^{(i)}}$ represents a colour space vector describing the
$i$-loop amplitude. The dependence on both renormalisation scale $\mu$ and
renormalisation scheme is implicit.

We denote the squared amplitude summed over spins and colours by
\begin{equation}
\braket{\cm}{\cm} = \A(s,t,u),
\end{equation}
where the Mandelstam variables are given by
\begin{equation}
s = (p_1+p_2)^2, \qquad t = (p_2+p_3)^2, \qquad u = (p_1+p_3)^2.
\end{equation}
For the physical processes, the spin and colour averaged 
amplitudes are related to $\A$ by
\begin{eqnarray}
\overline{\sum |{\cal M}({q + \bar{q} \to  \bar{q}^\prime + q^\prime } )|^2} &=&
\frac{1}{4N^2}~\A(s,t,u) \\
\overline{\sum |{\cal M}({q + q^\prime \to  q + q^\prime})|^2} &=&
\frac{1}{4N^2}~\A(u,t,s) \\
\overline{\sum |{\cal M}({q + \bar{q}^\prime \to  \bar{q}^\prime}+ q)|^2} &=& \frac{1}{4N^2}~\A(t,s,u) \\
\overline{\sum |{\cal M}({\bar{q} + \bar{q}^\prime \to  \bar{q} +
\bar{q}^\prime})|^2} &=& \frac{1}{4N^2}~\A(u,t,s).
\end{eqnarray}

The summed and squared amplitude has the perturbative expansion
\begin{equation}
\A(s,t,u) = 16\pi^2\as^2 \left[
 \A^4(s,t,u)+\left(\frac{\as}{2\pi}\right) \A^6(s,t,u)
 +\left(\frac{\as}{2\pi}\right)^2 \A^8(s,t,u) +
\O{\as^{3}}\right].
\end{equation}
In terms of the amplitudes
\begin{eqnarray}
\A^4(s,t,u) &=& \braket{\cm^{(0)}}{\cm^{(0)}} \equiv 2(N^2-1) \left(\frac{t^2+u^2}{s^2} - \epsilon
\right),\\
\A^6(s,t,u) &=& \left(
\braket{\cm^{(0)}}{\cm^{(1)}}+\braket{\cm^{(1)}}{\cm^{(0)}}\right),\\
\A^8(s,t,u) &=& \left( \braket{\cm^{(1)}}{\cm^{(1)}} +
\braket{\cm^{(0)}}{\cm^{(2)}}+\braket{\cm^{(2)}}{\cm^{(0)}}\right).
\end{eqnarray}
Expressions for $\A^6$ are given in Ref.~\cite{ES} using dimensional
regularisation to isolate the infrared and ultraviolet singularities.  

Here we concentrate on the next-to-next-to-leading order contribution $\A^8$ and
in particular the interference of the two-loop and tree graphs.

\section{Method}
\label{sec:method}

Massless two-loop integrals for $2 \to 2$ scattering can be described  
in terms of a basis set of scalar {\em master} integrals. 
The simpler massless master integrals comprise the trivial topologies
of single scale integrals which can be written as products of Gamma functions: 
\begin{eqnarray*}
{\rm Sunset}(s) &=& \SUNC{s}\\
{\rm Glass}(s) &=&   \GLASS{s} \\
{\rm Tri}(s) &=&  \TRI{s}
\end{eqnarray*}
the less trivial non-planar triangle graph~\cite{xtri},
$$
{\rm Xtri}(s) = \Xtri{s}
$$
and two scale integrals that are related to the one-loop box
graphs~\cite{AGO2,AGO3},
\begin{eqnarray*}
{\rm Abox}(s,t) &=&
\ABOX{s}{t} \\
{\rm Cbox}(s,t) &=& \CBOX{s}{t}. 
\end{eqnarray*}
The planar double box~\cite{planarA} and  non-planar double box~\cite{nonplanarA}
\begin{eqnarray*}
{\rm Pbox_1}(s,t) &=& \Pboxa{s}{t} \\
{\rm Xbox_1}(s,t) &=& \Xboxa{s}{t} 
\end{eqnarray*} 
involve multiple Mellin-Barnes integrals and 
are much more complicated to evaluate as series expansions in $\epsilon$.
Expressions for these integrals 
valid through to $\O{\epsilon^0}$ are given in~\cite{planarA} and
\cite{nonplanarA} respectively.

It turns out that for the two latter topologies, integrals involving loop momenta in the numerator 
cannot be entirely reduced in terms of the simpler integrals mentioned above and an
additional master integral is required in each case.
Reference~\cite{planarB} describes the procedure for reducing the tensor integrals down to a basis
involving the planar box integral 
$$
{\rm Pbox_2}(s,t) = \Pboxc{s}{t}, 
$$ 
where the  blob on the middle propagator represents an additional power of that propagator,
and provides a series expansion for ${\rm Pbox}_2$ to $\O{\epsilon^0}$.  However, as was
pointed out in~\cite{bastei}, knowledge of ${\rm Pbox_1}$ and ${\rm Pbox_2}$ to
$\O{\epsilon^0}$ is not sufficient to determine all tensor loop integrals to the same order. 
A better basis involves the tensor integral,
$$
{\rm Pbox_3}(s,t) = \Pboxb{s}{t}, 
$$ 
where  
\begin{picture}(1,1)
\thicklines
\put(0.5,0.2){\circle{0.5}}
\put(0.5,0.2){\makebox(0,0)[c]{$1$}}
\end{picture}
represents the planar box integral with one irreducible numerator associated with the left
loop. Symmetry of the integral
ensures that,
$$
\Pboxb{s}{t} \equiv \Pboxd{s}{t}.
$$
Series expansions for ${\rm Pbox_3}$ are relatively compact and straightforward to 
obtain and
are detailed in~\cite{bastei3,bastei2}.
${\rm Pbox_2}$ can therefore be eliminated in favor of ${\rm Pbox_3}$.
We note that this choice is not unique.   Bern et al.~\cite{BDG} choose to use
the ${\rm Pbox_1}$
and ${\rm Pbox_2}$ basis, but with the integrals evaluated in $D=6-2\epsilon$
dimensions where they are both infrared and ultraviolet finite. 
 
Similarly, the tensor reduction of the non-planar box integrals~\cite{nonplanarB} also
requires a second master integral,  
$$
{\rm Xbox_2}(s,t) =\Xboxb{s}{t},
$$
where the blob again denotes an additional power of the propagator. 
For the non-planar graphs there are no complications as in the planar case and 
all tensors to $\O{\epsilon^0}$ may be described in terms of the series expansions of ${\rm
Xbox_1}$ and ${\rm Xbox_2}$ through to 
$\O{\epsilon^0}$~\cite{nonplanarA,nonplanarB}.

In general tensor integrals are associated with scalar integrals in higher dimension and
with higher powers of propagators.    This connection can straightforwardly be achieved
using the Schwinger parameter form of the integral and is detailed in~\cite{AGO3} where
explicit expressions for generic two-loop integrals with up to four powers of loop momenta
in the numerator are given\footnote{A method to reduce tensor integrals constructing
differential operators that change the powers of the propagators as well as the dimension of
the integral was presented in Ref.~\cite{Tar}.}.  Systematic application of the
integration-by-parts (IBP) identities~\cite{IBP} and Lorentz invariance (LI) identities
\cite{diffeq} is sufficient to reduce these higher-dimension, higher-power integrals to
master integrals in $D=4-2\epsilon$.   Some topologies that occur in Feynman diagrams such
as the pentabox~\cite{AGO3} are immediately simplified using the IBP identities and collapse
to combinations of master integrals.   However, the tensor integrals directly associated
with the master integrals usually require more care. Explicit identities relevant for the
tensor integrals  of the ${\rm Abox}$ and ${\rm Cbox}$  topologies are given in~\cite{AGO3},
for   ${\rm Pbox_1}$ and ${\rm Pbox_2}$ integrals in~\cite{planarB}  while those for the
${\rm Xtri}$, ${\rm Xbox_1}$ and ${\rm Xbox_2}$ integrals are detailed in
\cite{nonplanarB}.  Using these identities, we have constructed MAPLE and FORM programs to
rewrite two-loop tensor integrals for massless $2 \to 2$ scattering directly in terms of the
basis set of master integrals.

The one-loop integrals are much better known.   There are only two master integrals,
the scalar bubble graph,
$$
{\rm Bub}(s) = \BUB{s},
$$
and the one-loop scalar box graph,
$$
{\rm Box}(s,t) = \BOX{s}{t}.
$$
We treat the tensor integrals in the same way as the two-loop integrals: shifting both
dimension and powers of propagators and then using IBP to rewrite the integrals as
combinations of ${\rm Bub}$ and ${\rm Box}$.  We note that this is not a unique
choice for the master integrals.   The one-loop bubble
graph is proportional to the one-loop triangle graph with one off-shell leg.
Another common choice is to replace the one-loop box in $D=4-2\epsilon$ by the
finite one-loop box in $D=6-2\epsilon$, ${\rm Box}^6$.

The general procedure for computing the amplitudes is therefore as follows.
First the two-loop Feynman diagrams are generated using {\tt QGRAF}~\cite{QGRAF}.
We then project by tree level, perform the summation over
colours and spins and trace over the Dirac matrices in $D$ dimensions using
conventional dimensional regularisation.
It is then straightforward to identify the scalar and tensor integrals present 
and replace
them with combinations of master integrals using the 
tensor reduction of two-loop integrals described in
\cite{planarB,nonplanarB,AGO3} based on integration-by-parts~\cite{IBP} and 
Lorentz invariance~\cite{diffeq} identities.   The final result is 
a combination of master integrals in $D=4-2\epsilon$ which can be substituted 
for the expansions in $\epsilon$ given in~\cite{planarA,nonplanarA,planarB,nonplanarB,
AGO2,AGO3,bastei3,bastei2}.  

\section{Results}
\label{sec:results}
In this section, we give explicit formulae for the $\epsilon$-expansion of the
two-loop contribution to the next-to-next-to-leading order term $\A^8(s,t,u)$.  
To distinguish between
the genuine two-loop contribution
$\braket{\cm^{(0)}}{\cm^{(2)}}+\braket{\cm^{(2)}}{\cm^{(0)}}$ and the squared
one-loop part $\braket{\cm^{(1)}}{\cm^{(1)}}$, we decompose $\A^8$ as
\begin{equation}
\A^8 = \A^{8~(2\times 0)} + \A^{8~(1\times 1)}.
\end{equation}
The one-loop-square contribution $\A^{8~(1\times 1)}$ is vital in determining 
$\A^8$ but is relatively straightforward to
obtain. 
For the remainder of this paper we concentrate on the technically more
complicated two-loop contribution $\A^{8~(2\times 0)}$.

We divide the two-loop contributions into two classes: those that multiply
poles in the dimensional regularisation parameter $\ep$ and those that are finite
as $\ep \to 0$
\beq
\A^{8~(2\times 0)}(s,t,u)
 = \Poles+\Finite.
\eeq 
$\Poles$ contains both infrared singularities and ultraviolet divergences.   The
latter are removed by renormalisation, while the former must be analytically
canceled by the infrared singularities occurring in radiative processes of the
same order.  The structure of these infrared divergences has been widely
studied and, as has been demonstrated by Catani~\cite{catani}, can be largely
predicted. 

\subsection{Infrared pole structure}

In the notation of Section~\ref{sec:notation}, the universal infrared divergences present in a one-loop amplitude are given by
the factorization formulae 
\beq
\label{eq:m1}
\ket{\cm^{(1)}}  = {\bom I}^{(1)}(\ep) \, \ket{\cm^{(0)}} + 
                   \ket{\cm^{(1)\, {\rm fin}}},
\eeq
where $\ket{\cm^{(1)\, {\rm fin}}}$ is finite as $\epsilon \to 0$ and the singular dependence is
determined by the colour-charge operator 
${\bom I}^{(1)}(\ep)$ that acts on the tree-level colour vector
$\ket{\cm^{(0)}}$.   
For the $n$ parton process we sum over all possible
colour antennae with colour operators $\TTT ij$ acting on the 
state  $\ket{\cm^{(0)}}$ to obtain
\beq
\label{eq:I1}
{\bom I}^{(1)}(\ep) = \frac{1}{2}\frac{e^{\ep \gamma}}{\Gamma(1-\ep)}
\sum_{i=1}^n \sum_{i\neq j}^n \TTT ij       \( \frac{1}{\ep^2}+ \frac{\gamma_i}{\ep}\)
\left( \frac{\mu^2 e^{-i\lambda_{ij}\pi}}{2p_i.p_j}\right)^{\epsilon}     \\
\eeq
and where $\lambda_{ij} = -1$ if $i$ and $j$ are both incoming or outgoing partons and
$\lambda_{ij}= 0$ otherwise and the constants
$\gamma_i$ are given by
\beq
\gamma_q = \gamma_{\bar q} = \frac{3}{2}, \qquad \gamma_g =
\frac{\beta_0}{\CA}.
\eeq

Similarly at the two-loop level there is a factorisation of the infrared singularities
\beq
\label{eq:m2}
\ket{\cm^{(2)}} = {\bom I}^{(1)}(\ep) \, \ket{\cm^{(1)}} +
                    {\bom I}^{(2)}(\ep) \, \ket{\cm^{(0)}} +\ket{\cm^{(2)\, {\rm fin}}} 
\eeq
where now
\beqn
\label{eq:I2}
{\bom I}^{(2)}(\ep)
&=&  - \frac{1}{2} {\bom I}^{(1)}(\ep)
\left( {\bom I}^{(1)}(\ep) + \frac{2\beta_0}{\ep} \right)
+e^{-\ep \gamma } \frac{ \Gamma(1-2\ep)}{\Gamma(1-\ep)} 
\left(\frac{\beta_0}{\ep} + K \right) {\bom I}^{(1)}(2\ep) + 
{\bom H}^{(2)}(\ep)\nonumber\\
\eeqn
where the constant $K$ is
\beq
K = \left( \frac{67}{18} - \frac{\pi^2}{6} \right) \CA - \frac{10}{9} T_R
\NF.
\eeq
The function ${\bom H}^{(2)}$ contains only single poles and is process dependent.
For the case of the quark form factor (in the \MSbar\  scheme)
it is given by
\beq
{\bom H}^{(2)}(\ep) = \frac{1}{4\epsilon} \frac{e^{\ep \gamma}}{\Gamma(1-\ep)} 
\left( \frac{\mu^2 e^{-i\lambda_{12}\pi}}{2p_1.p_2}\right)^{2\epsilon}
H^{(2)},
\eeq
with
\beq
H^{(2)} =  \left [\frac{1}{4}\gamma_{(1)}+3\CF K + \frac{5}{2}\zeta_2\beta_0\CF -
\frac{28}{9}\beta_0 \CF - \left(\frac{16}{9}-7\zeta_3\right)\CF\CA \right ]
\eeq
where $\zeta_n$ is the Riemann Zeta function, $\zeta_2 = \pi^2/6$, 
$\zeta_3 = 1.202056\ldots$  
and
\beq
\gamma_{(1)}=\(-3+24\zeta_2-48\zeta_3\)\CF^2
+\left(-\frac{17}{3}-\frac{88}{3}\zeta_2+24\zeta_3\right)\CF\CA
+\left(\frac{4}{3}+\frac{32}{3}\zeta_2\right)\CF T_R\NF.
\eeq
We expect that in the four-quark two loop amplitude, we might obtain contributions from
${\bom H}^{(2)}$ for each of the six colour antennae.   

Applying the formalism to the case at hand, we find that the pole structure of the two-loop
amplitude interfered with tree level has the following structure
\beqn
\label{eq:poles}
\Poles = 2 \Re \Biggl[ &&  \frac{1}{2}\bra{\cm^{(0)}} {\bom I}^{(1)}(\ep){\bom I}^{(1)}(\ep) \ket{\cm^{(0)}}
  -\frac{\beta_0}{\epsilon}  
\,\bra{\cm^{(0)}} {\bom I}^{(1)}(\ep)  \ket{\cm^{(0)}}
 \nonumber\\
&& 
+\,\bra{\cm^{(0)}} {\bom I}^{(1)}(\ep)  \ket{\cm^{(1)\, {\rm fin}}}
 \nonumber\\
&& 
+
e^{-\ep \gamma } \frac{ \Gamma(1-2\ep)}{\Gamma(1-\ep)} 
\left(\frac{\beta_0}{\epsilon} + K\right)
\bra{\cm^{(0)}} {\bom I}^{(1)}(2\ep) \ket{\cm^{(0)}}\nonumber \\
&&+ \, \bra{\cm^{(0)}}{\bom H}^{(2)}(\ep)\ket{\cm^{(0)}} \Biggr].
\eeqn
The colour algebra is straightforward and we find
\beqn
\lefteqn{\bra{\cm^{(0)}}{\bom I}^{(1)}(\ep)\ket{\cm^{(0)}} =
\braket{\cm^{(0)}}{\cm^{(0)}}}\nonumber\\
&&\times\frac{e^{\ep \gamma}}{\Gamma(1-\ep)}
 \( \frac{1}{\ep^2} + \frac{3}{2 \ep}\) 
\Bigg[
\frac{1}{N} \fs 
-\frac{2}{N} \fu 
-\frac{N^2-2}{N} \ft
\Bigg]  
\\ \nonumber \\\nonumber \\
\lefteqn{\bra{\cm^{(0)}}{\bom I}^{(1)}(\ep){\bom I}^{(1)}(\ep)\ket{\cm^{(0)}} 
=\braket{\cm^{(0)}}{\cm^{(0)}}}\nonumber\\
&&\times\frac{e^{2\ep \gamma}}{\Gamma(1-\ep)^2}
 \( \frac{1}{\ep^2} + \frac{3}{2 \ep}\)^2 \Bigg[
 \frac{N^4-3 N^2+3}{N^2} \ftd+ \frac{N^2+3}{N^2} \fud
\nonumber\\
&&{} \hspace{1cm}-2 \frac{N^2-2}{N^2}\fs \ft  +2
\frac{N^2-3}{N^2} \ft \fu 
\nonumber\\
&&{}\hspace{1cm}- \frac{4}{N^2}\fs \fu + \frac{1}{N^2} \fsd\Bigg]
\\ \nonumber \\\nonumber \\
\label{eq:I1M1}
\lefteqn{\bra{\cm^{(0)}}{\bom I}^{(1)}(\ep)\ket{\cm^{(1)\, {\rm fin}}} 
=\frac{e^{\ep \gamma}}{\Gamma(1-\ep)}
 \( \frac{1}{\ep^2} + \frac{3}{2 \ep}\) }\nonumber\\
&& \times \left\{ \left[{1\over N}\fs-\,{2\over N}\fu-{{N^2-2}\over
N}\,\ft\right]\,\F_1(s,t,u)\right.\nonumber \\
&& \hspace{1cm}+ \left.\Biggr[{1\over N}\fu -{1 \over N}\ft \Biggl]\,
(N^2-1)\,\F_2(s,t,u)
\right\}
\eeqn
and
\beqn
\label{eq:htwo}
\lefteqn{\bra{\cm^{(0)}}{\bom H}^{(2)}(\ep)\ket{\cm^{(0)}} = \braket{\cm^{(0)}}{\cm^{(0)}} }\nonumber \\ 
&&\hspace{1cm}\times\frac{e^{\ep \gamma}}{2\,\ep\,\Gamma(1-\ep)} H^{(2)} 
\Bigg[\fsd + \ftd - \fud 
\Bigg]\,,
\eeqn
where the square bracket in 
Eq.~(\ref{eq:htwo}) is a guess 
simply motivated by summing over the antennae present
in the quark-quark scattering process
and on
dimensional grounds.
Different choices only affect the finite remainder.

The functions $\F_1$ and $\F_2$ appearing in Eq.~(\ref{eq:I1M1}) 
are finite functions and are obtained from projection of
${\bom I}^{(1)}$ onto the one-loop
amplitude.  We find
\begin{eqnarray}
\lefteqn{\F_1(s,t,u) = \frac{N^2-1}{2N}
\left[ \(N^2-2\)f(s,t,u)+2f(s,u,t)\right]}\nonumber \\
&&-\frac{1}{2\ep (3-2\ep)}
\left[\frac{N^2-1}{N}\(6-7\ep-2\ep^2\)-\frac{1}{N}\left(
10\ep^2-4\ep^3\right)
\right]{\rm Bub}(s)
\braket{\cm^{(0)}}{\cm^{(0)}}\nonumber \\
&& -  
\frac{e^{\ep\gamma}}{\Gamma(1-\ep)}
\left(\frac{1}{\ep^2}+\frac{3}{2\ep}\right)  \left[\frac{1}{N}\fs
-\frac{2}{N}\fu
-\frac{N^2-2}{N}\ft \right] \braket{\cm^{(0)}}{\cm^{(0)}}\nonumber
\\
&&-
\beta_0 \left[
\frac{1}{\ep} 
-\frac{3(1-\ep)}{(3-2\ep)}{\rm Bub}(s)\right] \braket{\cm^{(0)}}{\cm^{(0)}}
\\ \nonumber \\
&&
\lefteqn{\F_2(s,t,u)= \frac{N^2-1}{2N}
\left[f(s,t,u)-f(s,u,t)\right]}\nonumber\\
&& -  
\frac{e^{\ep\gamma}}{\Gamma(1-\ep)}
\left(\frac{1}{\ep^2}+\frac{3}{2\ep}\right)  \left[\frac{1}{N}\fu-\frac{1}{N}\ft
\right] \braket{\cm^{(0)}}{\cm^{(0)}}
\eeqn
where the function $f(s,t,u)$ is written in terms of the 
one-loop box graph in $D=6-2\ep$ and the one-loop bubble graph in $D=4-2\ep$
\beqn
f(s,t,u)&=&
\frac{4(u^2+t^2)-2\epsilon(3ut+6t^2+5u^2)-\epsilon^2 s(7t+5u)}
{s^2} \left[ \frac{{\rm Bub}(s)-{\rm Bub}(t)}{\ep}\right]
\nonumber\\
&&+u\,(1-{2}\,{\epsilon})
\frac{6t^2+2u^2-3 \epsilon s^2}{ s^2} \, {\rm Box}^6(s,t).
\end{eqnarray}
These expressions are valid in all kinematic regions.  However, to evaluate the pole
structure in a particular region, they must be expanded as a series 
in $\epsilon$. We note that in Eq.~(\ref{eq:poles}), these functions are
multiplied by poles in $\epsilon$ and must therefore be expanded through to
$\O{\epsilon^2}$. In the physical region $ u < 0$, $t < 0$, $\Bfin$ has no imaginary part 
and is given by~\cite{BDG}
\begin{eqnarray}
\Bfin &=& \frac{ e^{\ep\gamma}
\Gamma  \left(  1+\epsilon \right)  \Gamma  
\left( 1-\epsilon \right) ^2 
 }{ 2s\Gamma  \left( 1-2 \epsilon  \right)   \left( 1-2 \epsilon  \right) }
 \left(\frac{\mu^2}{s}\right)^{\ep}
  \Biggl [
 \frac{1}{2}\(\(\lnx-\lny\)^2+\pi^2 \)\nonumber \\
&& 
 +2\ep \left(
 \Licx-\lnx\Libx-\frac{1}{3}\lnx^3-\frac{\pi^2}{2}\lnx \right)
\nonumber \\
&& 
-2\ep^2\Biggl(
\Lidx+\lny\Licx-\frac{1}{2}\lnx^2\Libx-\frac{1}{8}\lnx^4-\frac{1}{6}\lnx^3\lny+\frac{1}{4}\lnx^2\lny^2\nonumber
\\
&&-\frac{\pi^2}{4}\lnx^2-\frac{\pi^2}{3}\lnx\lny-\frac{\pi^4}{45}\Biggr)
+ ( u \leftrightarrow t) \Biggr ] + \O{\ep^3},
\end{eqnarray}
where $x = -t/s$, $\lnx=\log(x)$ and $\lny=\log(1-x)$ and the 
polylogarithms ${\rm Li}_n(z)$ are defined by
\begin{eqnarray}
 {\rm Li}_n(z) &=& \int_0^z \frac{dt}{t} {\rm Li}_{n-1}(t) \qquad {\rm ~for~}
 n=2,3,4\\
 {\rm Li}_2(z) &=& -\int_0^z \frac{dt}{t} \log(1-t).
\end{eqnarray} 
Analytic continuation to other kinematic regions is obtained 
using the inversion formulae for the arguments of 
the polylogarithms (see for example~\cite{AGO3}) when $x > 1$
\begin{eqnarray}
\lib(x + i0) &=& -\lib\left(\frac{1}{x}\right) -\frac{1}{2}
 \log^2 (x) +\frac{\pi^2}{3} + i \pi \log (x) \nonumber \\
\lic(x + i0) &=& \lic\left(\frac{1}{x}\right) -\frac{1}{6}\log^3(x) +
\frac{\pi^2}{3}  \log(x) +  \frac{i\pi}{2} \log^2(x) \nonumber \\ 
\lid(x + i0)  &=& - \lid\left(\frac{1}{x}\right)
         -  \frac{1}{24}\log^4(x)  + \frac{\pi^2}{6} \log^2(x)
        + \frac{\pi^4}{45} +
        \frac{i\pi}{6}\log^3(x).
\end{eqnarray}
Finally, the one-loop bubble integral in $D=4-2 \epsilon$ dimensions 
is given by
\begin{equation} 
 \Bubl =\frac{ e^{\ep\gamma}\Gamma  \left(  1+\epsilon \right)  \Gamma  \left( 1-\epsilon \right) ^2 
 }{ \Gamma  \left( 2-2 \epsilon  \right)  \epsilon   } \fs.
\end{equation}

Our explicit Feynman diagram reproduces the anticipated pole structure exactly
and provides a very stringent check on the calculation.  We therefore construct
the finite remainder by subtracting Eq.~(\ref{eq:poles}) from the full result. 
 
\subsection{Finite contributions}

In this subsection, we give explicit expressions for the finite two-loop
contribution to $\A^8$, $\Finite$, which is given by
\beq
\Finite = 2 \Re \braket{\cm^{(0)}}{\cm^{(2)\, {\rm fin}}}.
\eeq
For high energy hadron-hadron collisions, we probe all parton-parton
scattering processes simultaneously.  
We therefore need to be able to evaluate the
finite parts in the $s$-, $t$- and $u$-channels
corresponding to the processes
\begin{eqnarray*}
q + \bar{q} &\to&  \bar{q}^\prime + q^\prime\\
q + \bar{q}^\prime &\to&  \bar{q}^\prime +q\\
q + q^\prime &\to&  q + q^\prime,
\end{eqnarray*}
respectively.   
In principle, the analytic expressions for different channels 
are related by crossing
symmetry.   However, the ${\rm Xbox}$ has cuts in all three channels
yielding complex parts in all physical regions.   The analytic
continuation is therefore rather involved and prone to error.   We
therefore choose to give expressions 
describing $\A^8(s,t,u)$, $\A^8(t,s,u)$ and
$\A^8(u,t,s)$ which are directly 
valid in the physical region, $s > 0$ and
$u, t < 0$, and are given in terms of logarithms and polylogarithms that
have no imaginary parts.

In general the expansions of the two-loop master integrals
\cite{planarA,nonplanarA,planarB,nonplanarB,AGO3,bastei3,bastei2}
contain the 
generalised polylogarithms of Nielsen
\begin{equation}
{\rm S}_{n,p}(x) = \frac{(-1)^{n+p-1}}{(n-1)!\, p!} \int_0^1 dt \, 
  \frac{\log^{n-1}(t) \log^{p}(1-xt)}{t}, \quad \quad \quad n,p \ge 1, \ \
  x\le 1
\end{equation}
where the level is $n+p$.
Keeping terms up to $\O{\epsilon}$ corresponds to 
probing level 4 so that only polylogarithms with  $n+p \leq 4$ occur.
For $p=1$ we find the usual polylogarithms
\begin{equation}
{\rm S}_{n-1,1}(z) \equiv  {\rm Li}_n(z).
\end{equation}
A basis set of 6 polylogarithms (one with $n+p=2$, two with $n+p=3$ and three
with $n+p = 4$ is sufficient to describe a function of level 4.
At level 4, we choose to eliminate the $S_{22}$, $S_{13}$ and $S_{12}$ functions
using the standard polylogarithm identities~\cite{kolbig} 
and retain the polylogarithms with arguments $x$, $1-x$ and
$(x-1)/x$, where
\begin{equation}
\label{eq:xydef}
x = -\frac{t}{s}, \qquad y = -\frac{u}{s} = 1-x, \qquad -\frac{u}{t} =
\frac{x-1}{x}.
\end{equation}
For convenience, we also introduce the following logarithms
\begin{equation}
\lnx = \log\left(\frac{-t}{s}\right),
\qquad \lny = \log\left(\frac{-u}{s}\right),
\qquad \Ls = \log\left(\frac{s}{\mu^2}\right)
\end{equation}
where $\mu$ is the renormalisation scale.
The common choice $\mu^2 = s$ corresponds to setting $\Ls = 0$.

For each channel, we choose to present our results by grouping terms 
according to the
power of the number of colours $N$ and the number 
of light quarks $\NF$ so that in channel $c$ 
\begin{equation}
\label{eq:zi}
\Finite_c =
 2 \biggl ( N^2-1\biggr )
 \left(N^2 A_c + B_c  + \frac{1}{N^2} C_c
+ N \NF D_c   +\frac{\NF}{N} E_c  + \NF^2 F_c\right).
\end{equation}

\subsubsection{The $s$-channel process $q \bar q \to \bar q^\prime q^\prime$}
\label{subsec:sex}
We first give expressions for the $s$-channel annihilation process,
$q \bar q \to \bar q^\prime q^\prime$.
We find that
\begin{eqnarray}
A_s&=&{}\Biggl [{2}\,{\Lidx}+{}\Biggl ({}-{2}\,{\lnx}-{11\over 3}\Biggr ){}\,{\Licx}+{}\Biggl ({\lnx^2}+{11\over 3}\,{\lnx}-{2\over 3}\,{\pi^2}\Biggr ){}\,{\Libx}\nonumber \\
&&+{121\over 18}\,{\Ls^2} +{}\Biggl ({}-{11\over 3}\,{\lnx^2}+{11}\,{\lnx}-{296\over 27}\Biggr ){}\,{\Ls}+{1\over 6}\,{\lnx^4}+{}\Biggl ({1\over 3}\,{\lny}-{49\over 18}\Biggr ){}\,{\lnx^3} \nonumber \\
&&+{}\Biggl ({11\over 6}\,{\lny}-{5\over 6}\,{\pi^2}+{197\over 18}\Biggr ){}\,{\lnx^2}+{}\Biggl ({}-{2\over 3}\,{\lny}\,{\pi^2}-{47\over 18}\,{\pi^2}+{6}\,{\zeta_3}-{95\over 24}\Biggr ){}\,{\lnx} \nonumber \\
&&+{}\Biggl ({11\over 24}\,{\pi^2}-{7}\,{\zeta_3}-{409\over 216}\Biggr ){}\,{\lny} +{113\over 720}\,{\pi^4}-{7\over 6}\,{\pi^2}+{197\over 36}\,{\zeta_3}+{23213\over 2592}\Biggr ]{}\,{\tspus}\nonumber \\
&&+{}\Biggl [{}-{3}\,{\Lidy}+{6}\,{\Lidx}-{3}\,{\Lidz}+{}\Biggl ({}-{2}\,{\lnx}-{7\over 2}\Biggr ){}\,{\Licx} \nonumber \\
&&+{3}\,{\lnx}\,{\Licy}+{}\Biggl ({1\over 2}\,{\lnx^2}+{7\over 2}\,{\lnx}+{1\over 2}\,{\pi^2}\Biggr ){}\,{\Libx}+{}\Biggl ({}-{11\over 6}\,{\lnx^2}+{11\over 6}\,{\lnx}\Biggr ){}\,{\Ls} \nonumber \\
&&+{}\Biggl ({1\over 2}\,{\lny}\,{\pi^2}-{13\over 9}\,{\pi^2}-{\zeta_3}-{32\over 9}\Biggr ){}\,{\lnx}+{}\Biggl ({7\over 4}\,{\lny}-{3\over 4}\,{\pi^2}+{44\over 9}\Biggr ){}\,{\lnx^2} \nonumber \\
&&+{}\Biggl ({1\over 2}\,{\lny}-{49\over 36}\Biggr ){}\,{\lnx^3}-{7\over 120}\,{\pi^4}+{47\over 36}\,{\pi^2}+{2}\,{\zeta_3}\Biggr ]{}\,{\tsmus}+ \Biggl [{}{3}\,{\lnx^2} \Biggr ]{}\,{\tou} \nonumber \\
&&+{3}\,{\Lidy}-{3}\,{\Lidx}+{3}\,{\Lidz}-{3}\,{\lnx}\,{\Licy}-{5\over 2}\,{\Licx} \nonumber \\
&&+{}\Biggl ({5\over 2}\,{\lnx}-{1\over 2}\,{\pi^2}\Biggr ){}\,{\Libx}-{11\over 6}\,{\lnx}\,{\Ls}+{1\over 8}\,{\lnx^4}+{}\Biggl ({}-{1\over 2}\,{\lny}+{1\over 3}\Biggr ){}\,{\lnx^3} \nonumber \\
&&+{}\Biggl ({5\over 4}\,{\lny}+{1\over 4}\,{\pi^2}+{1\over 6}\Biggr ){}\,{\lnx^2}+{}\Biggl ({}-{1\over 2}\,{\lny}\,{\pi^2}-{7\over 6}\,{\pi^2}+{3}\,{\zeta_3}+{32\over 9}\Biggr ){}\,{\lnx} \nonumber \\
&&+{1\over 40}\,{\pi^4}-{11\over 36}\,{\pi^2}+{4}\,{\zeta_3}
\end{eqnarray}
\begin{eqnarray}
B_s&&={}\Biggl [{}-{6}\,{\Lidx}-{22\over 3}\,{\Licy}+{}\Biggl ({}-{3}\,{\lnx^2}-{22\over 3}\,{\lnx}-{22\over 3}\,{\lny}+{2}\,{\pi^2}\Biggr ){}\,{\Libx}\nonumber \\
&&+{}\Biggl ({6}\,{\lnx}+{22\over 3}\Biggr ){}\,{\Licx}+{}\Biggl ({22\over 3}\,{\lnx^2}-{22}\,{\lnx}-{22\over 3}\,{\lny^2}+{22}\,{\lny}-{88\over 3}\Biggr ){}\,{\Ls} \nonumber \\
&&-{1\over 2}\,{\lnx^4}+{}\Biggl ({}-{\lny}+{125\over 18}\Biggr ){}\,{\lnx^3}+{}\Biggl ({1\over 2}\,{\lny^2}-{31\over 6}\,{\lny}+{3}\,{\pi^2}-{743\over 36}\Biggr ){}\,{\lnx^2} \nonumber \\
&&+{}\Biggl ({}-{31\over 6}\,{\lny^2}+{}\Biggl ({}-{4\over 3}\,{\pi^2}+{9\over 2}\Biggr ){}\,{\lny}+{307\over 72}\,{\pi^2}+{\zeta_3}-{49\over 27}\Biggr ){}\,{\lnx} \nonumber \\
&&+{1\over 4}\,{\lny^4}-{71\over 18}\,{\lny^3}+{}\Biggl ({}-{2\over 3}\,{\pi^2}+{689\over 36}\Biggr ){}\,{\lny^2}+{}\Biggl ({}-{73\over 24}\,{\pi^2}-{\zeta_3}-{275\over 27}\Biggr ){}\,{\lny} \nonumber \\
&&+{79\over 720}\,{\pi^4}-{55\over 72}\,{\pi^2}-{443\over 36}\,{\zeta_3}+{30659\over 648}\Biggr ]{}\,{\tspus} \nonumber \\
&&+{}\Biggl [{}-{12}\,{\Lidy}+{3}\,{\Lidx}-{8}\,{\Lidz}+{}\Biggl ({2}\,{\lny}+{8}\Biggr ){}\,{\Licy} \nonumber \\
&&+{}\Biggl ({}-{3\over 2}\,{\lnx^2}+{}\Biggl ({}-{8}\,{\lny}-{11\over 2}\Biggr ){}\,{\lnx}+{8}\,{\lny}-{4\over 3}\,{\pi^2}\Biggr ){}\,{\Libx}-{12}\,{\lny}\,{\lnx}\,{\Liby} \nonumber \\
&&+{}\Biggl ({4}\,{\lnx}-{12}\,{\lny}+{11\over 2}\Biggr ){}\,{\Licx}+{}\Biggl ({11\over 3}\,{\lnx^2}-{11\over 3}\,{\lnx}+{11\over 3}\,{\lny^2}-{11\over 3}\,{\lny}\Biggr ){}\,{\Ls} \nonumber \\
&&-{17\over 24}\,{\lnx^4}+{}\Biggl ({\lny}+{131\over 36}\Biggr ){}\,{\lnx^3}+{}\Biggl ({}-{25\over 2}\,{\lny^2}-{15\over 4}\,{\lny}+{13\over 12}\,{\pi^2}-{289\over 36}\Biggr ){}\,{\lnx^2} \nonumber \\
&&+{}\Biggl ({1\over 3}\,{\lny^3}+{5}\,{\lny^2}+{5\over 3}\,{\lny}\,{\pi^2}+{89\over 36}\,{\pi^2}+{37\over 9}\Biggr ){}\,{\lnx}-{1\over 6}\,{\lny^4}+{17\over 9}\,{\lny^3} \nonumber \\
&&+{}\Biggl ({7\over 12}\,{\pi^2}-{361\over 36}\Biggr ){}\,{\lny^2}+{}\Biggl ({59\over 36}\,{\pi^2}+{6}\,{\zeta_3}+{64\over 9}\Biggr ){}\,{\lny}-{1\over 20}\,{\pi^4}-{44\over 9}\,{\pi^2}-{9}\,{\zeta_3}\Biggr ]{}\,{\tsmus} \nonumber \\
&&\Biggl [{}-{7}\,{\lnx^2}\Biggr ]{}\,{\tou}+\Biggl [{}{5}\,{\lny^2}\Biggr ]{}\,{\uot}-{12}\,{\Lidy}+{12}\,{\Lidx}-{12}\,{\Lidz} \nonumber \\
&&+{}\Biggl ({6}\,{\lnx}-{6}\Biggr ){}\,{\Licy}+{}\Biggl ({}-{6}\,{\lny}+{9\over 2}\Biggr ){}\,{\Licx}-{6}\,{\lny}\,{\lnx}\,{\Liby} \nonumber \\
&&+{}\Biggl ({}\Biggl ({}-{6}\,{\lny}-{9\over 2}\Biggr ){}\,{\lnx}-{6}\,{\lny}+{2}\,{\pi^2}\Biggr ){}\,{\Libx}+{}\Biggl ({11\over 3}\,{\lnx}-{11\over 3}\,{\lny}\Biggr ){}\,{\Ls} \nonumber \\
&&-{1\over 2}\,{\lnx^4}+{}\Biggl ({2}\,{\lny}-{5\over 6}\Biggr ){}\,{\lnx^3}+{}\Biggl ({}-{15\over 2}\,{\lny^2}-{2}\,{\lny}-{\pi^2}+{17\over 12}\Biggr ){}\,{\lnx^2} \nonumber \\
&&+{}\Biggl ({}-{11\over 4}\,{\lny^2}+{}\Biggl ({3}\,{\pi^2}+{1\over 2}\Biggr ){}\,{\lny}+{25\over 12}\,{\pi^2}-{6}\,{\zeta_3}-{37\over 9}\Biggr ){}\,{\lnx}+{1\over 4}\,{\lny^3} \nonumber \\
&&+{7\over 12}\,{\lny^2}+{}\Biggl ({}-{5\over 4}\,{\pi^2}+{6}\,{\zeta_3}+{64\over 9}\Biggr ){}\,{\lny}-{17\over 60}\,{\pi^4}-{2\over 3}\,{\pi^2}+{5}\,{\zeta_3}
\end{eqnarray}
\begin{eqnarray}
C_s&=&{}\Biggl [{16}\,{\Lidy}+{8}\,{\Lidx}-{16}\,{\lny}\,{\Licy}-{8}\,{\lnx}\,{\Licx}+{}\Biggl ({4}\,{\lnx^2}+{8\over 3}\,{\pi^2}\Biggr ){}\,{\Libx} \nonumber \\
&&+{8}\,{\lny^2}\,{\Liby}+{5\over 12}\,{\lnx^4}+{}\Biggl ({4\over 3}\,{\lny}-{9\over 2}\Biggr ){}\,{\lnx^3}+{}\Biggl ({}-{3\over 2}\,{\lny^2}+{9\over 2}\,{\lny}-{11\over 3}\,{\pi^2}+{1\over 4}\Biggr ){}\,{\lnx^2} \nonumber \\
&&+{}\Biggl ({8\over 3}\,{\lny^3}+{9\over 2}\,{\lny^2}+{}\Biggl ({22\over 3}\,{\pi^2}-{27\over 2}\Biggr ){}\,{\lny}+{1\over 2}\,{\pi^2}-{6}\,{\zeta_3}+{189\over 8}\Biggr ){}\,{\lnx} \nonumber \\
&&+{1\over 12}\,{\lny^4}-{9\over 2}\,{\lny^3}+{}\Biggl ({}-{7\over 3}\,{\pi^2}+{65\over 4}\Biggr ){}\,{\lny^2}+{}\Biggl ({}-{1\over 2}\,{\pi^2}+{6}\,{\zeta_3}-{189\over 8}\Biggr ){}\,{\lny} \nonumber \\
&&-{49\over 60}\,{\pi^4}+{29\over 24}\,{\pi^2}-{15\over 2}\,{\zeta_3}+{511\over 32}\Biggr ]{}\,{\tspus} \nonumber \\
&&+{}\Biggl [{12}\,{\Lidy}-{24}\,{\Lidx}+{24}\,{\Lidz}+{}\Biggl ({}-{18}\,{\lnx}+{10}\,{\lny}-{2}\Biggr ){}\,{\Licy} \nonumber \\
&&+{}\Biggl ({}-{2}\,{\lnx}+{18}\,{\lny}+{4}\Biggr ){}\,{\Licx}+{}\Biggl ({2}\,{\lnx^2}+{}\Biggl ({6}\,{\lny}-{4}\Biggr ){}\,{\lnx}-{2}\,{\lny}+{4}\,{\pi^2}\Biggr ){}\,{\Libx} \nonumber \\
&&+{}\Biggl ({18}\,{\lnx}\,{\lny}-{4}\,{\lny^2}\Biggr ){}\,{\Liby}+{4\over 3}\,{\lnx^4}+{}\Biggl ({}-{3}\,{\lny}-{8\over 3}\Biggr ){}\,{\lnx^3} \nonumber \\
&&+{}\Biggl ({15}\,{\lny^2}+{\lny}+{1\over 12}\,{\pi^2}-{15\over 4}\Biggr ){}\,{\lnx^2}+{}\Biggl ({}-{\lny^3}-{4}\,{\lny^2}-{2}\,{\lny}\,{\pi^2}+{11\over 12}\,{\pi^2}+{8}\,{\zeta_3}+{6}\Biggr ){}\,{\lnx} \nonumber \\
&&-{1\over 6}\,{\lny^4}+{7\over 3}\,{\lny^3}+{}\Biggl ({7\over 12}\,{\pi^2}-{9\over 4}\Biggr ){}\,{\lny^2}+{}\Biggl ({3\over 4}\,{\pi^2}-{16}\,{\zeta_3}+{6}\Biggr ){}\,{\lny} \nonumber \\
&&+{1\over 30}\,{\pi^4}-{4\over 3}\,{\pi^2}+{4}\,{\zeta_3}\Biggr ]{}\,{\tsmus}+\Biggl [{}{3}\,{\lnx^2}\Biggr ]{}\,{\tou}+\Biggl [{}{3}\,{\lny^2}\Biggr ]{}\,{\uot} \nonumber \\
&&+{4}\,{\Licy}+{2}\,{\Licx}+{}\Biggl ({}-{2}\,{\lnx}+{4}\,{\lny}\Biggr ){}\,{\Libx}+{3\over 4}\,{\lnx^3}+{}\Biggl ({}-{7\over 4}\,{\lny}-{15\over 4}\Biggr ){}\,{\lnx^2} \nonumber \\
&&+{}\Biggl ({5\over 4}\,{\lny^2}-{3\over 2}\,{\lny}+{5\over 12}\,{\pi^2}-{6}\Biggr ){}\,{\lnx}+{7\over 4}\,{\lny^3}+{9\over 4}\,{\lny^2}+{}\Biggl ({}-{19\over 12}\,{\pi^2}+{6}\Biggr ){}\,{\lny} \nonumber \\
&&+{\pi^2}-{12}\,{\zeta_3}
\end{eqnarray}
\begin{eqnarray}
D_s&=&{}\Biggl [{2\over 3}\,{\Licx}-{2\over 3}\,{\lnx}\,{\Libx}-{22\over 9}\,{\Ls^2}+{}\Biggl ({}-{2}\,{\lnx}+{2\over 3}\,{\lnx^2}+{389\over 54}\Biggr ){}\,{\Ls} \nonumber \\
&&+{2\over 9}\,{\lnx^3}+{}\Biggl ({}-{29\over 18}-{1\over 3}\,{\lny}\Biggr ){}\,{\lnx^2}+{}\Biggl ({10\over 9}\,{\pi^2}+{11\over 6}\Biggr ){}\,{\lnx} \nonumber \\
&&+{}\Biggl ({}-{1\over 12}\,{\pi^2}+{25\over 54}\Biggr ){}\,{\lny}-{455\over 54}+{41\over 36}\,{\pi^2}-{49\over 18}\,{\zeta_3}\Biggr ]{}\,{\tspus} \nonumber \\
&&+{}\Biggl [{}\Biggl ({1\over 3}\,{\lnx^2}-{1\over 3}\,{\lnx}\Biggr ){}\,{\Ls}+{1\over 9}\,{\lnx^3}-{13\over 18}\,{\lnx^2}+{}\Biggl ({4\over 9}\,{\pi^2}+{8\over 9}\Biggr ){}\,{\lnx}-{2\over 9}\,{\pi^2}\Biggr ]{}\,{\tsmus} \nonumber \\
&&+{1\over 3}\,{\lnx}\,{\Ls}-{1\over 6}\,{\lnx^2}-{8\over 9}\,{\lnx}+{2\over 9}\,{\pi^2}
\end{eqnarray}
\begin{eqnarray}
\label{eq:es}
E_s&=&{}\Biggl [{4\over 3}\,{\Licy}-{4\over 3}\,{\Licx}+{}\Biggl ({4\over 3}\,{\lny}+{4\over 3}\,{\lnx}\Biggr ){}\,{\Libx}-{4\over 9}\,{\lnx^3}+{}\Biggl ({29\over 9}+{2\over 3}\,{\lny}\Biggr ){}\,{\lnx^2} \nonumber \\
&&+{}\Biggl ({4}\,{\lnx}-{4}\,{\lny}+{4\over 3}\,{\lny^2}+{29\over 6}-{4\over 3}\,{\lnx^2}\Biggr ){}\,{\Ls}+{}\Biggl ({}-{223\over 54}+{2\over 3}\,{\lny^2}-{77\over 36}\,{\pi^2}\Biggr ){}\,{\lnx} \nonumber \\
&&+{4\over 9}\,{\lny^3}-{29\over 9}\,{\lny^2}+{}\Biggl ({223\over 54}+{23\over 12}\,{\pi^2}\Biggr ){}\,{\lny}-{35\over 18}\,{\zeta_3}-{685\over 81}-{7\over 36}\,{\pi^2}\Biggr ]{}\,{\tspus} \nonumber \\
&&+{}\Biggl [{}\Biggl ({}-{2\over 3}\,{\lnx^2}+{2\over 3}\,{\lnx}+{2\over 3}\,{\lny}-{2\over 3}\,{\lny^2}\Biggr ){}\,{\Ls}-{2\over 9}\,{\lnx^3}+{13\over 9}\,{\lnx^2} \nonumber \\
&&+{}\Biggl ({}-{8\over 9}\,{\pi^2}-{16\over 9}\Biggr ){}\,{\lnx}-{2\over 9}\,{\lny^3}+{13\over 9}\,{\lny^2}+{}\Biggl ({}-{8\over 9}\,{\pi^2}-{16\over 9}\Biggr ){}\,{\lny}+{8\over 9}\,{\pi^2}\Biggr ]{}\,{\tsmus} \nonumber \\
&&+{}\Biggl ({}-{2\over 3}\,{\lnx}+{2\over 3}\,{\lny}\Biggr ){}\,{\Ls}-{16\over 9}\,{\lny}-{1\over 3}\,{\lny^2}+{16\over 9}\,{\lnx}+{1\over 3}\,{\lnx^2}
\end{eqnarray}
\begin{eqnarray}
\label{eq:fs}
F_s&=&{}\Biggl ({}-{20\over 27}\,{\Ls}+{50\over 81}-{2\over 9}\,{\pi^2}+{2\over 9}\,{\Ls^2}\Biggr ){}\,{\tspus}
\end{eqnarray}

We can check some of these results by comparing with the analytic
expressions presented in Ref.~\cite{BDG} for the QED process
$e^+e^-\to \mu^+\mu^-$.
Taking the QED limit corresponds to setting $\CA = 0$, $\CF = 1$, $T_R
= 1$ as well as setting the cubic Casimir $C_3 = (N^2-1)(N^2-2)/N^2 = 0$.
This means that we can directly compare $E_s (\propto \CF T_R\NF)$ 
and $F_s (\propto T_R^2\NF^2)$ but {\em not} $C_s$ which receives
contributions from both $C_3$ and $\CF^2$.
We see that (\ref{eq:es}) 
and (\ref{eq:fs}) agree with  Eqs.~(2.38) and (2.39) of~\cite{BDG} 
respectively. 
 
The other coefficients, $A_s$, $B_s$, $C_s$ and $D_s$ are new results.

\subsubsection{The $t$-channel process $q + \bar{q}^\prime \to  q +
\bar{q}^\prime$}
\label{subsec:tex}
The $t$-channel process,
$q + \bar{q}^\prime \to  q +
\bar{q}^\prime$
is fixed by $\A^8(t,s,u)$. 
We find that the finite two-loop contribution in the $t$-channel is given by Eq.~(\ref{eq:zi}) with 
\begin{eqnarray}
A_t&=&{ }\Biggl [{}-{2}\,{\Lidx}+{}\Biggl ({2}\,{\lnx}-{11\over 3}\Biggr ){}\,{\Licx}+{}\Biggl ({11\over 3}\,{\lnx}-{\lnx^2}+{2\over 3}\,{\pi^2}\Biggr ){}\,{\Libx} \nonumber \\
&&+{121\over 18}\,{\Ls^2}+{}\Biggl ({22\over 9}\,{\lnx}-{11\over 3}\,{\lnx^2}-{296\over 27}\Biggr ){}\,{\Ls}+{1\over 4}\,{\lnx^4}+{}\Biggl ({}-{14\over 9}-{1\over 3}\,{\lny}\Biggr ){}\,{\lnx^3} \nonumber \\
&&+{}\Biggl ({}-{7\over 6}\,{\pi^2}+{20\over 3}+{11\over 6}\,{\lny}\Biggr ){}\,{\lnx^2}+{}\Biggl ({\zeta_3}-{46\over 9}+{2\over 3}\,{\lny}\,{\pi^2}-{373\over 72}\,{\pi^2}\Biggr ){}\,{\lnx} \nonumber \\
&&+{}\Biggl ({}-{409\over 216}+{11\over 24}\,{\pi^2}-{7}\,{\zeta_3}\Biggr ){}\,{\lny}+{23213\over 2592}-{49\over 9}\,{\pi^2}+{197\over 36}\,{\zeta_3}-{1\over 48}\,{\pi^4}\Biggr ]{}\,{\sspus} \nonumber \\
&&+{}\Biggl [{}-{3}\,{\Lidz}-{3}\,{\Lidy}-{6}\,{\Lidx}+{}\Biggl ({5}\,{\lnx}-{7\over 2}\Biggr ){}\,{\Licx} \nonumber \\
&&+{3}\,{\lnx}\,{\Licy}+{}\Biggl ({}-{1\over 2}\,{\pi^2}-{1\over 2}\,{\lnx^2}+{7\over 2}\,{\lnx}\Biggr ){}\,{\Libx}+{}\Biggl ({}-{11\over 6}\,{\lnx}-{11\over 6}\,{\lnx^2}\Biggr ){}\,{\Ls} \nonumber \\
&&-{1\over 6}\,{\lnx^4}+{}\Biggl ({}-{19\over 18}+{\lny}\Biggr ){}\,{\lnx^3}+{}\Biggl ({}-{1\over 2}\,{\pi^2}+{55\over 18}+{7\over 4}\,{\lny}\Biggr ){}\,{\lnx^2} \nonumber \\
&&+{}\Biggl ({}-{1\over 2}\,{\lny}\,{\pi^2}-{20\over 9}\,{\pi^2}+{32\over 9}-{2}\,{\zeta_3}\Biggr ){}\,{\lnx}-{19\over 36}\,{\pi^2}+{29\over 120}\,{\pi^4}+{2}\,{\zeta_3}\Biggr ]{}\,{\ssmus} \nonumber \\
&&+\Biggl [{}{3}\,{\lnx^2}\Biggr ]{}\,{\sou}+{3}\,{\Lidz}+{3}\,{\Lidy}+{3}\,{\Lidx}+{}\Biggl ({}-{3}\,{\lnx}-{5\over 2}\Biggr ){}\,{\Licx} \nonumber \\
&&-{3}\,{\lnx}\,{\Licy}+{}\Biggl ({1\over 2}\,{\pi^2}+{5\over 2}\,{\lnx}\Biggr ){}\,{\Libx}+{11\over 6}\,{\lnx}\,{\Ls}+{1\over 4}\,{\lnx^4}+{}\Biggl ({}-{\lny}-{3\over 4}\Biggr ){}\,{\lnx^3} \nonumber \\
&&+{}\Biggl ({5\over 4}\,{\lny}+{2}\Biggr ){}\,{\lnx^2}+{}\Biggl ({1\over 2}\,{\lny}\,{\pi^2}-{32\over 9}+{7\over 6}\,{\pi^2}\Biggr ){}\,{\lnx}-{5\over 24}\,{\pi^4}+{55\over 36}\,{\pi^2}+{4}\,{\zeta_3}
\end{eqnarray}
\begin{eqnarray}
B_t&=&{ }\Biggl [{6}\,{\Lidx}+{}\Biggl ({}-{6}\,{\lnx}+{44\over 3}\Biggr ){}\,{\Licx}+{}\Biggl ({22\over 3}\,{\lny}-{44\over 3}\,{\lnx}-{2}\,{\pi^2}+{3}\,{\lnx^2}\Biggr ){}\,{\Libx} \nonumber \\
&&+{22\over 3}\,{\Licy}+{}\Biggl ({}-{88\over 3}-{22\over 3}\,{\lny^2}+{22}\,{\lny}-{22\over 3}\,{\pi^2}+{44\over 3}\,{\lnx}\,{\lny}\Biggr ){}\,{\Ls}-{\lny}\,{\lnx^3} \nonumber \\
&&+{}\Biggl ({}-{\lny^3}+{29\over 3}\,{\lny^2}+{}\Biggl ({}-{187\over 9}-{7\over 3}\,{\pi^2}\Biggr ){}\,{\lny}-{52\over 3}+{25\over 3}\,{\pi^2}\Biggr ){}\,{\lnx} \nonumber \\
&&+{}\Biggl ({4}\,{\pi^2}+{2}\,{\lny^2}+{3}-{16\over 3}\,{\lny}\Biggr ){}\,{\lnx^2}+{1\over 4}\,{\lny^4}-{71\over 18}\,{\lny^3}+{}\Biggl ({5\over 6}\,{\pi^2}+{689\over 36}\Biggr ){}\,{\lny^2} \nonumber \\
&&+{}\Biggl ({}-{407\over 72}\,{\pi^2}-{\zeta_3}-{275\over 27}\Biggr ){}\,{\lny}+{30659\over 648}-{77\over 720}\,{\pi^4}-{707\over 36}\,{\zeta_3}+{183\over 8}\,{\pi^2}\Biggr ]{}\,{\sspus} \nonumber \\
&&+{}\Biggl [{}-{12}\,{\Lidz}-{8}\,{\Lidy}-{3}\,{\Lidx}+{}\Biggl ({}-{14}\,{\lny}+{10}\,{\lnx}-{5\over 2}\Biggr ){}\,{\Licx} \nonumber \\
&&+{}\Biggl ({}-{8}-{2}\,{\lny}+{2}\,{\lnx}\Biggr ){}\,{\Licy}+{}\Biggl ({}-{5\over 2}\,{\lnx^2}+{}\Biggl ({4}\,{\lny}+{5\over 2}\Biggr ){}\,{\lnx}-{8}\,{\lny}-{8\over 3}\,{\pi^2}\Biggr ){}\,{\Libx} \nonumber \\
&&+{}\Biggl ({22\over 3}\,{\lnx^2}+{}\Biggl ({}-{22\over 3}\,{\lny}+{22\over 3}\Biggr ){}\,{\lnx}+{11\over 3}\,{\pi^2}-{11\over 3}\,{\lny}+{11\over 3}\,{\lny^2}\Biggr ){}\,{\Ls}-{5\over 12}\,{\lnx^4} \nonumber \\
&&+{}\Biggl ({73\over 18}+{\lny}\Biggr ){}\,{\lnx^3}+{}\Biggl ({}-{41\over 12}\,{\lny}-{3\over 2}\,{\lny^2}-{193\over 18}+{11\over 6}\,{\pi^2}\Biggr ){}\,{\lnx^2} \nonumber \\
&&+{}\Biggl ({1\over 3}\,{\lny^3}-{7}\,{\lny^2}+{}\Biggl ({7\over 6}\,{\pi^2}+{295\over 18}\Biggr ){}\,{\lny}-{101\over 9}-{8}\,{\zeta_3}+{92\over 9}\,{\pi^2}\Biggr ){}\,{\lnx}-{1\over 6}\,{\lny^4}+{17\over 9}\,{\lny^3} \nonumber \\
&&+{}\Biggl ({}-{5\over 12}\,{\pi^2}-{361\over 36}\Biggr ){}\,{\lny^2}+{}\Biggl ({64\over 9}+{167\over 36}\,{\pi^2}+{8}\,{\zeta_3}\Biggr ){}\,{\lny}-{\zeta_3}+{29\over 90}\,{\pi^4}-{91\over 12}\,{\pi^2}\Biggr ]{}\,{\ssmus} \nonumber \\
&&-\Biggl [{}{7}\,{\lnx^2}\Biggr ]{}\,{\sou}+{}\Biggl [{5}\,{\lny^2}-{10}\,{\lnx}\,{\lny}+{5}\,{\pi^2}+{5}\,{\lnx^2}\Biggr ]{}\,{\uos}-{12}\,{\Lidz} \nonumber \\
&&-{12}\,{\Lidy}-{12}\,{\Lidx}+{}\Biggl ({12}\,{\lnx}-{6}\,{\lny}+{21\over 2}\Biggr ){}\,{\Licx}+{}\Biggl ({6}+{6}\,{\lnx}\Biggr ){}\,{\Licy} \nonumber \\
&&+{}\Biggl ({}-{21\over 2}\,{\lnx}-{2}\,{\pi^2}+{6}\,{\lny}\Biggr ){}\,{\Libx}-{11\over 3}\,{\Ls}\,{\lny}-{1\over 2}\,{\lnx^4}+{}\Biggl ({}-{1\over 6}+{3}\,{\lny}\Biggr ){}\,{\lnx^3} \nonumber \\
&&+{}\Biggl ({5\over 2}+{1\over 2}\,{\pi^2}-{3\over 4}\,{\lny}-{3\over 2}\,{\lny^2}\Biggr ){}\,{\lnx^2}+{}\Biggl ({}-{16\over 3}\,{\lny}-{3}+{2}\,{\lny^2}-{29\over 6}\,{\pi^2}-{6}\,{\zeta_3}\Biggr ){}\,{\lnx} \nonumber \\
&&+{1\over 4}\,{\lny^3}+{7\over 12}\,{\lny^2}+{}\Biggl ({}-{2}\,{\pi^2}+{6}\,{\zeta_3}+{64\over 9}\Biggr ){}\,{\lny}+{5\over 12}\,{\pi^2}+{13\over 20}\,{\pi^4}-{\zeta_3}
\end{eqnarray}
\begin{eqnarray}
C_t&=&{ }\Biggl [{16}\,{\Lidz}+{}\Biggl ({16}\,{\lny}-{8}\,{\lnx}\Biggr ){}\,{\Licx}+{}\Biggl ({}-{16}\,{\lnx}+{16}\,{\lny}\Biggr ){}\,{\Licy} \nonumber \\
&&-{8}\,{\Lidx}+{}\Biggl ({4}\,{\lnx^2}+{8}\,{\lny^2}+{16\over 3}\,{\pi^2}-{16}\,{\lnx}\,{\lny}\Biggr ){}\,{\Libx}+{2\over 3}\,{\lnx^4}-{4\over 3}\,{\lny}\,{\lnx^3} \nonumber \\
&&+{}\Biggl ({3}-{5}\,{\lny^2}-{5\over 3}\,{\pi^2}\Biggr ){}\,{\lnx^2}+{}\Biggl ({5}\,{\lny^3}+{9}\,{\lny^2}+{}\Biggl ({7\over 3}\,{\pi^2}-{19}\Biggr ){}\,{\lny}+{16}\,{\zeta_3}-{9}\,{\pi^2}\Biggr ){}\,{\lnx} \nonumber \\
&&+{1\over 12}\,{\lny^4}-{9\over 2}\,{\lny^3}+{}\Biggl ({65\over 4}-{19\over 6}\,{\pi^2}\Biggr ){}\,{\lny^2}+{}\Biggl ({}-{189\over 8}-{10}\,{\zeta_3}-{5}\,{\pi^2}\Biggr ){}\,{\lny} \nonumber \\
&&+{\pi^4}-{15\over 2}\,{\zeta_3}+{511\over 32}+{95\over 24}\,{\pi^2}\Biggr ]{}\,{\sspus} \nonumber \\
&&+{}\Biggl [{12}\,{\Lidz}+{24}\,{\Lidy}+{24}\,{\Lidx}+{}\Biggl ({}-{24}\,{\lnx}+{8}\,{\lny}+{6}\Biggr ){}\,{\Licx} \nonumber \\
&&+{}\Biggl ({6}\,{\lnx^2}+{}\Biggl ({}-{6}-{4}\,{\lny}\Biggr ){}\,{\lnx}-{4}\,{\lny^2}+{4}\,{\pi^2}+{2}\,{\lny}\Biggr ){}\,{\Libx} \nonumber \\
&&+{}\Biggl ({}-{10}\,{\lny}-{8}\,{\lnx}+{2}\Biggr ){}\,{\Licy}+{5\over 6}\,{\lnx^4}+{}\Biggl ({2}\,{\lny}-{2}\,{\lny^2}-{7\over 3}\,{\pi^2}-{6}\Biggr ){}\,{\lnx^2} \nonumber \\
&&+{}\Biggl ({}-{2\over 3}\,{\lny}+{2\over 3}\Biggr ){}\,{\lnx^3}+{}\Biggl ({}-{7\over 3}\,{\lny^3}-{3}\,{\lny^2}+{}\Biggl ({9\over 2}+{19\over 6}\,{\pi^2}\Biggr ){}\,{\lny}+{16}\,{\zeta_3}-{19\over 3}\,{\pi^2}-{12}\Biggr ){}\,{\lnx} \nonumber \\
&&-{1\over 6}\,{\lny^4}+{7\over 3}\,{\lny^3}+{}\Biggl ({1\over 4}\,{\pi^2}-{9\over 4}\Biggr ){}\,{\lny^2}+{}\Biggl ({13\over 12}\,{\pi^2}+{6}-{6}\,{\zeta_3}\Biggr ){}\,{\lny} \nonumber \\
&&-{43\over 12}\,{\pi^2}-{7\over 4}\,{\pi^4}+{2}\,{\zeta_3}\Biggr ]{}\,{\ssmus}+\Biggl [{}{3}\,{\lnx^2}\Biggr ]{}\,{\sou}+{}\Biggl [{3}\,{\pi^2}-{6}\,{\lnx}\,{\lny}+{3}\,{\lny^2}+{3}\,{\lnx^2}\Biggr ]{}\,{\uos} \nonumber \\
&&-{2}\,{\Licx}-{4}\,{\Licy}+{}\Biggl ({2}\,{\lnx}-{4}\,{\lny}\Biggr ){}\,{\Libx}+{}\Biggl ({}-{3}+{2}\,{\lny}\Biggr ){}\,{\lnx^2} \nonumber \\
&&+{}\Biggl ({\pi^2}-{3}\,{\lny}-{13\over 2}\,{\lny^2}\Biggr ){}\,{\lnx}+{7\over 4}\,{\lny^3}+{9\over 4}\,{\lny^2}+{}\Biggl ({6}+{7\over 2}\,{\pi^2}\Biggr ){}\,{\lny}+{7\over 4}\,{\pi^2}-{8}\,{\zeta_3}
\end{eqnarray}
\begin{eqnarray}
D_t&=&{ }\Biggl [{2\over 3}\,{\Licx}-{2\over 3}\,{\lnx}\,{\Libx}-{22\over 9}\,{\Ls^2}+{}\Biggl ({}-{26\over 9}\,{\lnx}+{2\over 3}\,{\lnx^2}+{389\over 54}\Biggr ){}\,{\Ls} \nonumber \\
&&+{5\over 9}\,{\lnx^3}+{}\Biggl ({}-{37\over 18}-{1\over 3}\,{\lny}\Biggr ){}\,{\lnx^2}+{}\Biggl ({265\over 54}+{11\over 36}\,{\pi^2}\Biggr ){}\,{\lnx}+{}\Biggl ({25\over 54}-{1\over 12}\,{\pi^2}\Biggr ){}\,{\lny} \nonumber \\
&&-{49\over 18}\,{\zeta_3}-{455\over 54}+{25\over 36}\,{\pi^2}\Biggr ]{}\,{\sspus} \nonumber \\
&&+{}\Biggl [{}\Biggl ({1\over 3}\,{\lnx^2}+{1\over 3}\,{\lnx}\Biggr ){}\,{\Ls}+{2\over 9}\,{\lnx^3}-{7\over 18}\,{\lnx^2}+{}\Biggl ({}-{8\over 9}+{2\over 9}\,{\pi^2}\Biggr ){}\,{\lnx}+{1\over 9}\,{\pi^2}\Biggr ]{}\,{\ssmus} \nonumber \\
&&-{1\over 3}\,{\lnx}\,{\Ls}-{1\over 9}\,{\pi^2}+{8\over 9}\,{\lnx}-{1\over 2}\,{\lnx^2}
\end{eqnarray}
\begin{eqnarray}
E_t&=&{ }\Biggl [{}-{8\over 3}\,{\Licx}-{4\over 3}\,{\Licy}+{}\Biggl ({}-{4\over 3}\,{\lny}+{8\over 3}\,{\lnx}\Biggr ){}\,{\Libx}-{2\over 3}\,{\lnx^2}\,{\lny} \nonumber \\
&&+{}\Biggl ({}-{4}\,{\lny}+{4\over 3}\,{\pi^2}+{4\over 3}\,{\lny^2}+{29\over 6}-{8\over 3}\,{\lnx}\,{\lny}\Biggr ){}\,{\Ls}+{}\Biggl ({29\over 6}+{22\over 9}\,{\lny}-{2\over 3}\,{\lny^2}+{2\over 3}\,{\pi^2}\Biggr ){}\,{\lnx} \nonumber \\
&&+{4\over 9}\,{\lny^3}-{29\over 9}\,{\lny^2}+{}\Biggl ({223\over 54}+{37\over 36}\,{\pi^2}\Biggr ){}\,{\lny}-{41\over 12}\,{\pi^2}-{685\over 81}-{11\over 18}\,{\zeta_3}\Biggr ]{}\,{\sspus} \nonumber \\
&&+{}\Biggl [{}\Biggl ({}-{4\over 3}\,{\lnx^2}+{}\Biggl ({4\over 3}\,{\lny}-{4\over 3}\Biggr ){}\,{\lnx}-{2\over 3}\,{\pi^2}+{2\over 3}\,{\lny}-{2\over 3}\,{\lny^2}\Biggr ){}\,{\Ls}-{8\over 9}\,{\lnx^3} \nonumber \\
&&+{}\Biggl ({14\over 9}+{2\over 3}\,{\lny}\Biggr ){}\,{\lnx^2}+{}\Biggl ({}-{20\over 9}\,{\lny}-{8\over 9}\,{\pi^2}+{32\over 9}\Biggr ){}\,{\lnx}-{2\over 9}\,{\lny^3}+{13\over 9}\,{\lny^2} \nonumber \\
&&+{}\Biggl ({}-{16\over 9}-{2\over 9}\,{\pi^2}\Biggr ){}\,{\lny}+{\pi^2}\Biggr ]{}\,{\ssmus} \nonumber \\
&&+{4\over 3}\,{\lnx}\,{\lny}+{2\over 3}\,{\Ls}\,{\lny}-{1\over 3}\,{\lny^2}-{1\over 3}\,{\pi^2}-{16\over 9}\,{\lny}
\end{eqnarray}
\begin{eqnarray}
F_t&=&{ }\Biggl [{2\over 9}\,{\Ls^2}+{}\Biggl ({4\over
9}\,{\lnx}-{20\over 27}\Biggr ){}\,{\Ls}+{2\over 9}\,{\lnx^2}-{20\over
27}\,{\lnx}+{50\over 81}\Biggr ]{}\,{\sspus} 
\end{eqnarray}

\subsubsection{The $u$-channel process $q + q^\prime \to  q + q^\prime$}
\label{subsec:uex}
The $u$-channel process,
$q + q^\prime \to  q + q^\prime$ is determined by $A^8(u,t,s)$.
We find that the finite two-loop contribution in the $u$-channel is given by Eq.~(\ref{eq:zi}) with 
\begin{eqnarray} 
A_u&=&{ }\Biggl [{}-{2}\,{\Lidz}+{}\Biggl ({2}\,{\lnx}+{11\over 3}-{2}\,{\lny}\Biggr ){}\,{\Licx}
+{}\Biggl ({2}\,{\lnx}+{11\over 3}-{2}\,{\lny}\Biggr ){}\,{\Licy}
\nonumber \nonumber \\ &&
+{}\Biggl ({}-{\lnx^2}+{}\Biggl ({}-{11\over 3}+{2}\,{\lny}\Biggr ){}\,{\lnx}-{1\over 3}\,{\pi^2}+{11\over 3}\,{\lny}-{\lny^2}\Biggr ){}\,{\Libx}+{121\over 18}\,{\Ls^2}
\nonumber \nonumber \\ &&
+{}\Biggl ({}-{11\over 3}\,{\lnx^2}+{}\Biggl ({11}+{22\over 3}\,{\lny}\Biggr ){}\,{\lnx}-{11\over 3}\,{\pi^2}-{296\over 27}+{22\over 9}\,{\lny}-{11\over 3}\,{\lny^2}\Biggr ){}\,{\Ls}+{1\over 12}\,{\lnx^4}
\nonumber \nonumber \\ &&
+{}\Biggl ({}-{49\over 18}-{2\over 3}\,{\lny}\Biggr ){}\,{\lnx^3}+{}\Biggl ({2}\,{\lny^2}+{197\over 18}+{8\over 3}\,{\lny}\Biggr ){}\,{\lnx^2}+{}\Biggl ({}-{5\over 3}\,{\lny^3}
\nonumber \nonumber \\ &&
+{17\over 6}\,{\lny^2}+{}\Biggl ({}-{2\over 3}\,{\pi^2}-{98\over 9}\Biggr ){}\,{\lny}+{4}\,{\zeta_3}-{95\over 24}-{31\over 9}\,{\pi^2}\Biggr ){}\,{\lnx}+{1\over 4}\,{\lny^4}
\nonumber \nonumber \\ &&
-{14\over 9}\,{\lny^3}+{}\Biggl ({1\over 2}\,{\pi^2}+{20\over 3}\Biggr ){}\,{\lny^2}+{}\Biggl ({}-{46\over 9}-{25\over 8}\,{\pi^2}+{3}\,{\zeta_3}\Biggr ){}\,{\lny}+{17\over 144}\,{\pi^4}+{65\over 36}\,{\zeta_3}
\nonumber \nonumber \\ &&
+{11\over 2}\,{\pi^2}+{23213\over 2592}\Biggr ){}\,{\tspss}+{}\Biggl ({}-{6}\,{\Lidz}+{3}\,{\Lidx}+{3}\,{\Lidy}
\nonumber \nonumber \\ &&
+{}\Biggl ({2}\,{\lnx}-{2}\,{\lny}+{7\over 2}\Biggr ){}\,{\Licx}+{}\Biggl ({}-{5}\,{\lny}+{7\over 2}+{5}\,{\lnx}\Biggr ){}\,{\Licy}
\nonumber \nonumber \\ &&
+{}\Biggl ({}-{1\over 2}\,{\lnx^2}+{}\Biggl ({}-{7\over 2}+{\lny}\Biggr ){}\,{\lnx}-{\pi^2}+{7\over 2}\,{\lny}-{1\over 2}\,{\lny^2}\Biggr ){}\,{\Libx}
\nonumber \nonumber \\ &&
+{}\Biggl ({}-{11\over 6}\,{\lnx^2}+{}\Biggl ({11\over 6}+{11\over 3}\,{\lny}\Biggr ){}\,{\lnx}-{11\over 6}\,{\pi^2}-{11\over 6}\,{\lny}-{11\over 6}\,{\lny^2}\Biggr ){}\,{\Ls}-{1\over 8}\,{\lnx^4}
\nonumber \nonumber \\ &&
+{}\Biggl ({1\over 2}\,{\lny}-{49\over 36}\Biggr ){}\,{\lnx^3}+{}\Biggl ({}-{1\over 4}\,{\pi^2}+{1\over 2}\,{\lny}+{3\over 4}\,{\lny^2}+{44\over 9}\Biggr ){}\,{\lnx^2}+{}\Biggl ({}-{5\over 6}\,{\lny^3}
\nonumber \nonumber \\ &&
+{37\over 12}\,{\lny^2}+{}\Biggl ({}-{5\over 6}\,{\pi^2}-{143\over 18}\Biggr ){}\,{\lny}-{32\over 9}-{3}\,{\zeta_3}-{67\over 36}\,{\pi^2}\Biggr ){}\,{\lnx}-{1\over 24}\,{\lny^4}-{19\over 18}\,{\lny^3}
\nonumber \nonumber \\ &&
+{}\Biggl ({}-{5\over 12}\,{\pi^2}+{55\over 18}\Biggr ){}\,{\lny^2}+{}\Biggl ({}-{83\over 36}\,{\pi^2}+{3}\,{\zeta_3}+{32\over 9}\Biggr ){}\,{\lny}-{3\over 2}\,{\zeta_3}-{7\over 120}\,{\pi^4}+{157\over 36}\,{\pi^2}\Biggr ]{}\,{\tsmss}
\nonumber \nonumber \\ &&
+{}\Biggl [{}-{6}\,{\lnx}\,{\lny}+{3}\,{\lny^2}+{3}\,{\pi^2}+{3}\,{\lnx^2}\Biggr ]{}\,{\tos}
\nonumber \nonumber \\ &&
+{3}\,{\Lidz}-{3}\,{\Lidx}-{3}\,{\Lidy}+{5\over 2}\,{\Licx}+{}\Biggl ({5\over 2}-{3}\,{\lnx}+{3}\,{\lny}\Biggr ){}\,{\Licy}
\nonumber \nonumber \\ &&
+{}\Biggl ({5\over 2}\,{\lny}-{5\over 2}\,{\lnx}
+{1\over 2}\,{\pi^2}\Biggr ){}\,{\Libx}+{}\Biggl ({11\over 6}\,{\lny}-{11\over 6}\,{\lnx}\Biggr ){}\,{\Ls}+{1\over 8}\,{\lnx^4}
\nonumber \nonumber \\ &&
+{}\Biggl ({1\over 3}-{1\over 2}\,{\lny}\Biggr ){}\,{\lnx^3}+{}\Biggl ({1\over 6}+{1\over 4}\,{\pi^2}-{9\over 4}\,{\lny}\Biggr ){}\,{\lnx^2}+{}\Biggl ({7\over 2}\,{\lny^2}+{}\Biggl ({}-{13\over 6}+{1\over 2}\,{\pi^2}\Biggr ){}\,{\lny}
\nonumber \nonumber \\ &&
+{32\over 9}+{3}\,{\zeta_3}-{1\over 6}\,{\pi^2}\Biggr ){}\,{\lnx}+{1\over 8}\,{\lny^4}-{3\over 4}\,{\lny^3}+{}\Biggl ({2}+{1\over 2}\,{\pi^2}\Biggr ){}\,{\lny^2}
\nonumber \nonumber \\ &&
+{}\Biggl ({}-{3}\,{\zeta_3}-{32\over 9}-{3\over 2}\,{\pi^2}\Biggr ){}\,{\lny}+{3\over 2}\,{\zeta_3}+{61\over 36}\,{\pi^2}+{11\over 120}\,{\pi^4}
\end{eqnarray} 
\begin{eqnarray} 
B_u&=&{ }\Biggl [{6}\,{\Lidz}+{}\Biggl ({}-{22\over 3}+{6}\,{\lny}-{6}\,{\lnx}\Biggr ){}\,{\Licx}+{}\Biggl ({}-{44\over 3}-{6}\,{\lnx}+{6}\,{\lny}\Biggr ){}\,{\Licy}
\nonumber \nonumber \\ &&
+{}\Biggl ({3}\,{\lnx^2}+{}\Biggl ({}-{6}\,{\lny}+{22\over 3}\Biggr ){}\,{\lnx}+{3}\,{\lny^2}-{44\over 3}\,{\lny}+{\pi^2}\Biggr ){}\,{\Libx}+{}\Biggl ({22\over 3}\,{\lnx^2}
\nonumber \nonumber \\ &&
+{}\Biggl ({}-{22}-{44\over 3}\,{\lny}\Biggr ){}\,{\lnx}-{88\over 3}+{22\over 3}\,{\pi^2}\Biggr ){}\,{\Ls}-{1\over 4}\,{\lnx^4}+{}\Biggl ({2}\,{\lny}+{125\over 18}\Biggr ){}\,{\lnx^3}
\nonumber \nonumber \\ &&
+{}\Biggl ({}-{11\over 2}\,{\lny^2}-{743\over 36}-{25\over 3}\,{\lny}+{1\over 2}\,{\pi^2}\Biggr ){}\,{\lnx^2}+{}\Biggl ({4}\,{\lny^3}-{28\over 3}\,{\lny^2}+{}\Biggl ({133\over 9}+{7\over 3}\,{\pi^2}\Biggr ){}\,{\lny}
\nonumber \nonumber \\ &&
+{535\over 72}\,{\pi^2}+{7}\,{\zeta_3}-{49\over 27}\Biggr ){}\,{\lnx}+{}\Biggl ({}-{5\over 2}\,{\pi^2}+{3}\Biggr ){}\,{\lny^2}+{}\Biggl ({}-{52\over 3}+{1\over 9}\,{\pi^2}-{6}\,{\zeta_3}\Biggr ){}\,{\lny}
\nonumber \nonumber \\ &&
-{1217\over 72}\,{\pi^2}+{30659\over 648}-{437\over 720}\,{\pi^4}-{179\over 36}\,{\zeta_3}\Biggr ){}\,{\tspss}+{}\Biggl ({}-{3}\,{\Lidz}+{8}\,{\Lidx}
\nonumber \nonumber \\ &&
+{12}\,{\Lidy}+{}\Biggl ({}-{4}\,{\lnx}-{8}\,{\lny}-{11\over 2}\Biggr ){}\,{\Licx}+{}\Biggl ({5\over 2}-{10}\,{\lny}-{4}\,{\lnx}\Biggr ){}\,{\Licy}
\nonumber \nonumber \\ &&
+{}\Biggl ({3\over 2}\,{\lnx^2}+{}\Biggl ({11\over 2}+{\lny}\Biggr ){}\,{\lnx}-{5\over 2}\,{\lny^2}+{5\over 2}\,{\lny}-{7\over 6}\,{\pi^2}\Biggr ){}\,{\Libx}+{}\Biggl ({11\over 3}\,{\lnx^2}+{}\Biggl ({}-{11\over 3}
\nonumber \nonumber \\ &&
-{22\over 3}\,{\lny}\Biggr ){}\,{\lnx}+{11\over 3}\,{\pi^2}+{22\over 3}\,{\lny}+{22\over 3}\,{\lny^2}\Biggr ){}\,{\Ls}
-{1\over 2}\,{\lnx^4}+{}\Biggl ({131\over 36}+{7\over 3}\,{\lny}\Biggr ){}\,{\lnx^3}
\nonumber \nonumber \\ &&
+{}\Biggl (-{289\over 36}-{3\over 4}\,{\pi^2}-{15\over 4}\,{\lny^2}-{7\over 2}\,{\lny}\Biggr ){}\,{\lnx^2}+{}\Biggl ({}-{11\over 6}\,{\lny^3}+{13\over 12}\,{\lny^2}+{}\Biggl ({223\over 18}+{17\over 6}\,{\pi^2}\Biggr ){}\,{\lny}
\nonumber \nonumber \\ &&
+{73\over 18}\,{\pi^2}+{37\over 9}+{4}\,{\zeta_3}\Biggr ){}\,{\lnx}+{1\over 12}\,{\lny^4}+{73\over 18}\,{\lny^3}+{}\Biggl ({3\over 4}\,{\pi^2}-{193\over 18}\Biggr ){}\,{\lny^2}+{}\Biggl ({191\over 36}\,{\pi^2}
\nonumber \nonumber \\ &&
+{2}\,{\zeta_3}-{101\over 9}\Biggr ){}\,{\lny}-{67\over 12}\,{\pi^2}-{7\over 2}\,{\zeta_3}-{61\over 90}\,{\pi^4}\Biggr ]{}\,{\tsmss}
\nonumber \nonumber \\ &&
+{}\Biggl [{}-{7}\,{\pi^2}-{7}\,{\lnx^2}-{7}\,{\lny^2}
+{14}\,{\lnx}\,{\lny}\Biggr ]{}\,{\tos}+{5}\,{\sot}\,{\lny^2}-{12}\,{\Lidz}+{12}\,{\Lidx}
\nonumber \nonumber \\ &&
+{12}\,{\Lidy}+{}\Biggl ({}-{9\over 2}-{6}\,{\lny}\Biggr ){}\,{\Licx}+{}\Biggl ({6}\,{\lnx}-{21\over 2}-{12}\,{\lny}\Biggr ){}\,{\Licy}
\nonumber \nonumber \\ &&
+{}\Biggl ({}-{21\over 2}\,{\lny}+{9\over 2}\,{\lnx}-{2}\,{\pi^2}\Biggr ){}\,{\Libx}+{11\over 3}\,{\lnx}\,{\Ls}-{1\over 2}\,{\lnx^4}+{}\Biggl ({}-{5\over 6}+{2}\,{\lny}\Biggr ){}\,{\lnx^3}
\nonumber \nonumber \\ &&
+{}\Biggl ({}-{3\over 2}\,{\lny^2}+{17\over 12}+{9\over 2}\,{\lny}-{\pi^2}\Biggr ){}\,{\lnx^2}+{}\Biggl ({1\over 12}\,{\pi^2}-{37\over 4}\,{\lny^2}-{6}\,{\zeta_3}+{1\over 3}\,{\lny}-{37\over 9}-{\lny^3}\Biggr ){}\,{\lnx}
\nonumber \nonumber \\ &&
-{1\over 6}\,{\lny^3}+{}\Biggl ({5\over 2}-{3\over 2}\,{\pi^2}\Biggr ){}\,{\lny^2}+{}\Biggl ({65\over 12}\,{\pi^2}+{6}\,{\zeta_3}-{3}\Biggr ){}\,{\lny}+{5\over 4}\,{\pi^2}-{11\over 20}\,{\pi^4}+{19\over 2}\,{\zeta_3}
\end{eqnarray} 
\begin{eqnarray} 
C_u&=&{ }\Biggl [{}-{8}\,{\Lidz}-{16}\,{\Lidy}+{}\Biggl ({8}\,{\lnx}-{8}\,{\lny}\Biggr ){}\,{\Licx}
+{}\Biggl ({8}\,{\lnx}+{8}\,{\lny}\Biggr ){}\,{\Licy}
\nonumber \nonumber \\ &&
+{}\Biggl ({8}\,{\lnx}\,{\lny}+{4}\,{\lny^2}-{4}\,{\lnx^2}-{20\over 3}\,{\pi^2}\Biggr ){}\,{\Libx}+{1\over 12}\,{\lnx^4}+{}\Biggl ({}-{5\over 3}\,{\lny}-{9\over 2}\Biggr ){}\,{\lnx^3}
\nonumber \nonumber \\ &&
+{}\Biggl ({9}\,{\lny}+{1\over 4}+{5}\,{\lny^2}-{11\over 6}\,{\pi^2}\Biggr ){}\,{\lnx^2}+{}\Biggl ({8\over 3}\,{\lny^3}+{}\Biggl ({13}-{3}\,{\pi^2}\Biggr ){}\,{\lny}-{4}\,{\pi^2}+{189\over 8}-{14}\,{\zeta_3}\Biggr ){}\,{\lnx}
\nonumber \nonumber \\ &&
+{}\Biggl ({3}-{11\over 3}\,{\pi^2}\Biggr ){}\,{\lny^2}+{}\Biggl ({}-{9}\,{\pi^2}+{8}\,{\zeta_3}\Biggr ){}\,{\lny}+{9\over 5}\,{\pi^4}-{289\over 24}\,{\pi^2}-{15\over 2}\,{\zeta_3}+{511\over 32}\Biggr ]{}\,{\tspss}
\nonumber \nonumber \\ &&
+{}\Biggl [{24}\,{\Lidz}-{24}\,{\Lidx}-{12}\,{\Lidy}+{}\Biggl ({}-{4}+{16}\,{\lny}+{2}\,{\lnx}\Biggr ){}\,{\Licx}
\nonumber \nonumber \\ &&
+{}\Biggl ({}-{6}-{16}\,{\lnx}+{24}\,{\lny}\Biggr ){}\,{\Licy}+{}\Biggl ({}-{2}\,{\lnx^2}+{}\Biggl ({}-{8}\,{\lny}+{4}\Biggr ){}\,{\lnx}-{6}\,{\lny}
\nonumber \nonumber \\ &&
+{6}\,{\pi^2}+{6}\,{\lny^2}\Biggr ){}\,{\Libx}+{4\over 3}\,{\lnx^4}+{}\Biggl ({}-{19\over 3}\,{\lny}-{8\over 3}\Biggr ){}\,{\lnx^3}+{}\Biggl ({7}\,{\lny}+{2}\,{\lny^2}-{15\over 4}
\nonumber \nonumber \\ &&
+{25\over 12}\,{\pi^2}\Biggr ){}\,{\lnx^2}+{}\Biggl ({10\over 3}\,{\lny^3}-{10}\,{\lny^2}+{}\Biggl ({15\over 2}-{1\over 2}\,{\pi^2}\Biggr ){}\,{\lny}-{13\over 12}\,{\pi^2}+{6}\,{\zeta_3}+{6}\Biggr ){}\,{\lnx}
+{1\over 3}\,{\lny^4}
\nonumber \nonumber \\ &&
+{2\over 3}\,{\lny^3}+{}\Biggl ({5\over 3}\,{\pi^2}-{6}\Biggr ){}\,{\lny^2}+{}\Biggl ({20\over 3}\,{\pi^2}-{8}\,{\zeta_3}-{12}\Biggr ){}\,{\lny}-{61\over 12}\,{\pi^2}+{21\over 20}\,{\pi^4}+{8}\,{\zeta_3}\Biggr ]{}\,{\tsmss}
\nonumber \nonumber \\ &&
+{}\Biggl [{}-{6}\,{\lnx}\,{\lny}+{3}\,{\lny^2}+{3}\,{\pi^2}+{3}\,{\lnx^2}\Biggr ]{}\,{\tos}
\nonumber \nonumber \\ &&
+{3}\,{\sot}\,{\lny^2}-{2}\,{\Licx}+{2}\,{\Licy}+{}\Biggl ({2}\,{\lny}+{2}\,{\lnx}\Biggr ){}\,{\Libx}+{3\over 4}\,{\lnx^3}
\nonumber \nonumber \\ &&
+{}\Biggl ({}-{1\over 2}\,{\lny}-{15\over 4}\Biggr ){}\,{\lnx^2}+{}\Biggl ({9}\,{\lny}+{7\over 6}\,{\pi^2}-{6}\Biggr ){}\,{\lnx}+{8\over 3}\,{\lny}\,{\pi^2}-{17\over 4}\,{\pi^2}-{3}\,{\lny^2}-{10}\,{\zeta_3}
\nonumber \\\end{eqnarray} 
\begin{eqnarray} 
D_u&=&{ }\Biggl [{}-{2\over 3}\,{\Licx}-{2\over 3}\,{\Licy}+{}\Biggl ({2\over 3}\,{\lnx}-{2\over 3}\,{\lny}\Biggr ){}\,{\Libx}-{22\over 9}\,{\Ls^2} \nonumber \\ 
&&+{}\Biggl ({2\over 3}\,{\lnx^2}+{}\Biggl ({}-{2}-{4\over 3}\,{\lny}\Biggr ){}\,{\lnx}+{2\over 3}\,{\lny^2}-{26\over 9}\,{\lny}+{389\over 54}+{2\over 3}\,{\pi^2}\Biggr ){}\,{\Ls} \nonumber \\ 
&&+{2\over 9}\,{\lnx^3}+{}\Biggl ({1\over 3}\,{\lny}-{29\over 18}\Biggr ){}\,{\lnx^2}+{}\Biggl ({4\over 9}\,{\pi^2}-{4\over 3}\,{\lny^2}+{11\over 9}\,{\lny}+{11\over 6}\Biggr ){}\,{\lnx} \nonumber \\ 
&&+{5\over 9}\,{\lny^3}-{37\over 18}\,{\lny^2}+{}\Biggl ({3\over 4}\,{\pi^2}+{265\over 54}\Biggr ){}\,{\lny}-{455\over 54}-{11\over 12}\,{\pi^2}-{37\over 18}\,{\zeta_3}\Biggr ]{}\,{\tspss} \nonumber \\ 
&&+{}\Biggl [{}\Biggl ({1\over 3}\,{\lnx^2}+{}\Biggl ({}-{1\over 3}-{2\over 3}\,{\lny}\Biggr ){}\,{\lnx}+{1\over 3}\,{\lny}+{1\over 3}\,{\pi^2}+{1\over 3}\,{\lny^2}\Biggr ){}\,{\Ls} \nonumber \\ 
&&+{1\over 9}\,{\lnx^3}-{13\over 18}\,{\lnx^2}+{}\Biggl ({}-{1\over 3}\,{\lny^2}+{1\over 9}\,{\pi^2}+{8\over 9}+{10\over 9}\,{\lny}\Biggr ){}\,{\lnx}+{2\over 9}\,{\lny^3} \nonumber \\ 
&&-{7\over 18}\,{\lny^2}+{}\Biggl ({}-{8\over 9}+{2\over 9}\,{\pi^2}\Biggr ){}\,{\lny}-{11\over 18}\,{\pi^2}\Biggr ]{}\,{\tsmss} \nonumber \\ 
&&+{}\Biggl ({1\over 3}\,{\lnx}-{1\over 3}\,{\lny}\Biggr ){}\,{\Ls}-{1\over
6}\,{\lnx^2}+{}\Biggl ({}-{8\over 9}+{2\over 3}\,{\lny}\Biggr
){}\,{\lnx}-{5\over 18}\,{\pi^2}+{8\over 9}\,{\lny}-{1\over
2}\,{\lny^2}\nonumber \\
\end{eqnarray} 
\begin{eqnarray} 
E_u&=&{ }\Biggl [{4\over 3}\,{\Licx}+{8\over 3}\,{\Licy}+{}\Biggl ({}-{4\over 3}\,{\lnx}+{8\over 3}\,{\lny}\Biggr ){}\,{\Libx} \nonumber \\ 
&&+{}\Biggl ({}-{4\over 3}\,{\lnx^2}+{}\Biggl ({8\over 3}\,{\lny}+{4}\Biggr ){}\,{\lnx}-{4\over 3}\,{\pi^2}+{29\over 6}\Biggr ){}\,{\Ls}-{4\over 9}\,{\lnx^3}+{}\Biggl ({29\over 9}-{2\over 3}\,{\lny}\Biggr ){}\,{\lnx^2} \nonumber \\ 
&&+{}\Biggl ({10\over 3}\,{\lny^2}-{223\over 54}-{29\over 36}\,{\pi^2}-{22\over 9}\,{\lny}\Biggr ){}\,{\lnx}+{}\Biggl ({}-{10\over 9}\,{\pi^2}+{29\over 6}\Biggr ){}\,{\lny} \nonumber \\ 
&&-{685\over 81}-{59\over 18}\,{\zeta_3}+{109\over 36}\,{\pi^2}\Biggr ]{}\,{\tspss} \nonumber \\ 
&&+{}\Biggl [{}\Biggl ({}-{2\over 3}\,{\lnx^2}+{}\Biggl ({2\over 3}+{4\over 3}\,{\lny}\Biggr ){}\,{\lnx}-{4\over 3}\,{\lny}-{2\over 3}\,{\pi^2}-{4\over 3}\,{\lny^2}\Biggr ){}\,{\Ls}-{2\over 9}\,{\lnx^3}+{13\over 9}\,{\lnx^2} \nonumber \\ 
&&+{}\Biggl ({2\over 3}\,{\lny^2}-{20\over 9}\,{\lny}-{2\over 9}\,{\pi^2}-{16\over 9}\Biggr ){}\,{\lnx}-{8\over 9}\,{\lny^3}+{14\over 9}\,{\lny^2}+{}\Biggl ({32\over 9}-{8\over 9}\,{\pi^2}\Biggr ){}\,{\lny}+{\pi^2}\Biggr ]{}\,{\tsmss} \nonumber \\ 
&&-{2\over 3}\,{\lnx}\,{\Ls}+{1\over 3}\,{\lnx^2}+{}\Biggl ({}-{4\over 3}\,{\lny}+{16\over 9}\Biggr ){}\,{\lnx}+{1\over 3}\,{\pi^2}
\end{eqnarray} 
\begin{eqnarray} 
F_u&=&{ }\Biggl [{2\over 9}\,{\Ls^2}+{}\Biggl ({4\over 9}\,{\lny}-{20\over 27}\Biggr ){}\,{\Ls}+{2\over 9}\,{\lny^2}-{20\over 27}\,{\lny}+{50\over 81}\Biggr ]{}\,{\tspss}
\end{eqnarray}

\section{Summary}
\label{sec:conc}

In this paper we presented the two-loop QCD corrections to the scattering of
two distinct massless quarks.  Throughout, we have used conventional dimensional
regularisation and the \MSbar\  scheme to compute the interference of
the tree and two-loop graphs summed over spins and colours.  The pole structure
is given in Eq.~(\ref{eq:poles}) while expressions for the finite parts are
given for each of the $s$-, $t$- and $u$-channels in Secs.~\ref{subsec:sex},
\ref{subsec:tex} and \ref{subsec:uex} respectively.

The leading infrared singularity is $\O{1/\ep^4}$ and it is a very strong check
on the reliability of our calculation that the pole structure obtained by
computing the Feynman diagrams agrees with that anticipated by Catani through
to $\O{1/\ep}$.   For the finite $\NF/N$ and $\NF^2$ contributions, we agree
with prior QED calculations.   

These results form a crucial part of the next-to-next-to-leading order
predictions for jet cross sections in hadron-hadron collisions.   However, they
are only a part of the whole and must be combined with the  tree-level $2 \to
4$, the one-loop $2 \to 3$ as well as the square of the one-loop $2 \to 2$
processes to yield physical cross sections.    For the most part, the matrix
elements themselves are available in the literature.  Each of the contributions
is  divergent in the infrared limit and a systematic procedure for analytically
canceling the infrared divergences needs to be established for semi-inclusive 
jet cross sections.    Here again, some progress has been made by examining the
limits of tree-level matrix elements when two particles are unresolved
\cite{tc,ds} and the soft and collinear limits of one-loop amplitudes
\cite{sone,cone}.  The somewhat simpler case of $e^+e^- \to {\rm photon} + {\rm
jet}$ at next-to-leading order which involves the triple collinear limit of
tree-level matrix elements as well as the collinear limit of one-loop
amplitudes has already been studied~\cite{aude} and  indicates that the
technical problems are not insurmountable.   Another technical difficulty will
be to isolate the initial state singularities and correctly absorb them into
the parton density functions at next-to-next-to-leading order. Recent progress
towards the three-loop splitting functions~\cite{moms1,moms2,Gra1} together
with accurate parameterisations in $x$-space~\cite{NV,NVplb} suggest that the
factorisation can be achieved.   Modifications to the global fits to provide
parton density functions appropriate for next-to-next-to-leading order
calculations are already underway~\cite{MRS}.  In summary, we are therefore
confident that these problems will soon be overcome thereby enabling the
analytic cancellation of the infrared divergences and the construction of
numerical programs to provide next-to-next-to-leading order QCD estimates of
observable scattering cross sections.

\section*{Acknowledgements}

We thank  Adrian Signer and Bas Tausk for helpful suggestions and Stefano
Catani for clarifying the content of Ref.~\cite{catani}.

C.A. acknowledges the financial support of the Greek Government and
M.E.T. acknowledges financial support from CONACyT and the CVCP.  C.O. thanks
S.~Catani for useful discussions.  We gratefully acknowledge the support of
the British Council and German Academic Exchange Service under ARC project
1050.  This work was supported in part by the EU Fourth Framework Programme
`Training and Mobility of Researchers', Network `Quantum Chromodynamics and
the Deep Structure of Elementary Particles', contract FMRX-CT98-0194
(DG-12-MIHT).

\end{document}